\documentclass[aps,
	       prd,
	       nofootinbib,
	       preprintnumbers,
	       nobibnotes,
	       reprint,
	       twocolumn]{revtex4-1}

\usepackage{amssymb}
\usepackage{amsmath}
\usepackage{graphicx,subfigure}
\usepackage{longtable}
\usepackage{verbatim}
\usepackage{amsfonts}
\usepackage{hyperref}
\usepackage{color}

\newcommand{\eq}[1]{Eq.~(\ref{#1})}
\newcommand{\eqsand}[2]{Eqs.~(\ref{#1}) and (\ref{#2})}
\newcommand{\eqsto}[2]{Eqs.~(\ref{#1}) to (\ref{#2})}
\newcommand{\fig}[1]{Fig.~\ref{#1}}
\newcommand{\Dbar}{\,\overline{\!D}}
\newcommand{\dd}{\ensuremath{D\!-\!\Dbar{}\,}}
\newcommand{\ddm}{\dd\ mixing}
\newcommand{\ov}{\overline}

\newcommand{\be}{\begin{equation}}
\newcommand{\ee}{\end{equation}}
\newcommand{\bea}{\begin{eqnarray}}
\newcommand{\eea}{\end{eqnarray}}
\newcommand{\beq}{\begin{equation}}
\newcommand{\eeq}{\end{equation}}
\def\beqa{\begin{eqnarray}}
  \def\eeqa{\end{eqnarray}}
\newcommand{\bv}{\left(\begin{array}{c}}
\newcommand{\ev}{\end{array}\right)}

\def\lsim{\mathrel{\rlap{\lower4pt\hbox{\hskip1pt$\sim$}}
    \raise1pt\hbox{$<$}}}	  
\def\gsim{\mathrel{\rlap{\lower4pt\hbox{\hskip1pt$\sim$}}
    \raise1pt\hbox{$>$}}}	  

\newcommand{\bra}[1]{\left\langle{#1}\right\vert}
\newcommand{\ket}[1]{\left\vert{#1}\right\rangle}
\newcommand{\nn}{\nonumber}
\newcommand{\imag}{\mbox{Im}\,}
\newcommand{\real}{\mbox{Re}\,}
\newcommand{\ds}{\displaystyle}
\newcommand{\gev}{\ensuremath{\,}\mbox{GeV}}
\newcommand{\dm}{\ensuremath{\Delta M}}
\newcommand{\dg}{\ensuremath{\Delta \Gamma}}

\begin{document}

\preprint{TTP15-027}

\title{CP Violation in $\mathbf{D^0\rightarrow K_SK_S}$}
\author{Ulrich Nierste$^{\,a}$}
\email{ulrich.nierste@kit.edu}
\author{Stefan Schacht$^{\,a}$}
\email{stefan.schacht@kit.edu}
\affiliation{
$^{\,a}$ Institut f\"ur Theoretische Teilchenphysik, Karlsruher
  Institut f\"ur Technologie, 76128 Karlsruhe, Germany}

\vspace*{1cm}

\begin{abstract}
 The direct CP asymmetry  $a_{CP}^{\mathrm{dir}}(D^0\rightarrow
 K_SK_S)$ involves exchange diagrams which are induced at tree level 
 in the Standard Model. Since the corresponding topological amplitude
 $E_{KK}$ can be large, $D^0\rightarrow
 K_SK_S$ is a promising discovery channel for charm CP violation.  
 We estimate the penguin annihilation amplitude with a perturbative 
 calculation and extract the exchange amplitude $E_{KK}$ 
 from a global fit to $D$ branching ratios. Our results are further used 
 to predict the size of mixing-induced CP violation. We  
 obtain $\vert a_{CP}^{\mathrm{dir}}(D^0\rightarrow
 K_SK_S)\vert \leq 1.1\%$ (95\% C.L.).  The same bound applies to
 the nonuniversal part of the phase between the \ddm\ and decay
 amplitudes. If future data exceed our predictions, this will
 point to new physics or an enhancement of the penguin annihilation
 amplitude by QCD dynamics. We briefly discuss the implications of these
 possibilities for other 
 CP asymmetries.
\end{abstract}

\maketitle

\section{Introduction}
While direct CP violation (CPV) is well established in the down-quark
sector \cite{Fanti:1999nm,Lai:2001ki,Batley:2002gn,AlaviHarati:1999xp,
  AlaviHarati:2002ye, Lin:2008zzaa, Duh:2012ie, Aubert:2007mj, Lees:2012mma, 
  Abulencia:2006psa, Aaltonen:2011qt}, CPV has not yet been 
observed in the decays of up-type quarks.  For the discussion of CPV in
some singly Cabibbo-suppressed $D$ decay it is convenient to decompose
the decay amplitude $ \mathcal{A}$ as
\begin{align}
\mathcal{A} &= \lambda_{sd} \mathcal{A}_{sd} -
\frac{\lambda_b}{2} \mathcal{A}_b. \label{eq:asdb}
\end{align}
Here $\lambda_q\equiv V_{cq}^*V_{uq}$ and $\lambda_{sd} =
(\lambda_s-\lambda_d)/2$ comprise the elements $V_{ij}$ of the
Cabibbo-Kobayashi-Maskawa (CKM) matrix. In the limit $\lambda_b=0$ all
direct and mixing-induced CP asymmetries vanish in the Standard Model
(SM). The suppression factor $\imag \frac{\lambda_b}{\lambda_{sd}}\sim
-6\cdot 10^{-4}$ makes the discovery of CKM-induced CPV challenging. At
the same time this parametric suppression renders CP asymmetries in
charm decays highly sensitive to physics beyond the SM.

In this paper we study the decay $D^0\to K_S K_S$. For this decay mode
$\mathcal{A}_{sd}$ vanishes in the limit of exact
SU(3)$_F$ symmetry~\cite{Buccella:1994nf,Bhattacharya:2009ps,Brod:2011re,Hiller:2012xm}, so that the
branching ratio is suppressed. However, $\mathcal{A}_b$ does not vanish in this
limit and we expect $|\mathcal{A}_b/\mathcal{A}_{sd}|$ to be
large. Therefore CP asymmetries in $D^0\to K_SK_S$ may be enhanced to an
observable level, even if the Kobayashi-Maskawa phase is the only source
of CPV in charm decays~\cite{Brod:2011re,Hiller:2012xm}.  Moreover, a
special feature of $D^0\to K_S K_S$ is the interference of the decays
$c\ov u \to \ov s s$ and $c\ov u \to \ov d d$, both of which involve the
tree-level exchange of a $W$ boson (exchange topology $E$, see
\fig{fig:su3limit-penguin-annihilation}). This interference term gives a
contribution to $\mathcal{A}_b$ owing to
$\lambda_d+\lambda_s=-\lambda_b$.  That is, contrary to the widely
studied decays $D\to \pi^+\pi^-,\pi^0\pi^0,K^+K^-$, no penguin diagrams
are needed for nonzero direct or mixing-induced CP
asymmetries. Moreover, the exchange diagram $E$ is enhanced by a large
Wilson coefficient.
These properties make $D^0\to K_S K_S$ an interesting discovery channel
for CPV in the charm system.

In this paper we calculate the allowed ranges for the direct and
mixing-induced CP asymmetries in $D^0\to K_S K_S$, using the results of
our global analysis in Ref.~\cite{Muller:2015lua}. There are two
ingredients which we cannot extract from this analysis: the first one is
the penguin-annihilation amplitude $PA$ (see
\fig{fig:su3limit-penguin-annihilation}), which we estimate with the
help of a perturbative calculation. The other undetermined quantity is a
strong phase $\delta$, whose value, however, can be determined from the
data once both the direct and mixing-induced CP asymmetries are
measured.  The actual size of $\delta$ is not crucial for the potential
to discover charm CPV in $D^0\to K_S K_S$: depending {on} whether
$|\sin\delta|$ is large or small either the direct or mixing-induced CP
asymmetry will be large.

Our paper is organized as follows: in
  Section~\ref{sec:preliminaries} {we derive handy formulae}
  for direct and {mixing-induced} CP asymmetries in terms of
  $\mathcal{A}_{sd}$ and $\mathcal{A}_b$.  In
  Section~\ref{sec:topological} we {relate the CPV observables to} 
  topological amplitudes. {Subsequently we} estimate the penguin
  annihilation {contribution, which cannot be extracted from 
  a global fit to current data, with perturbative methods} in 
  Section~\ref{sec:estimate}.   
  {In Section~\ref{sec:pheno} we present our phenomenological
  analysis.} Finally, we conclude.

\section{Preliminaries \label{sec:preliminaries}}
In this section we collect the formulae for the CP asymmetries.
We write 
\begin{align} 
\mathcal{A} (D^0\rightarrow K_SK_S) = -\frac{1}{\sqrt{2}}\,
\mathcal{A} (D^0\rightarrow \bar{K}^0 K^0)\, ,
\label{eq:bose}
\end{align}
accommodating the Bose symmetrization of the two $K_S$'s with the factor of
$1/\sqrt{2}$.  Here we identify $K_S=(K^0-\ov K{}^0)/\sqrt2$ and assume that
the effects of kaon CPV are properly subtracted from CP asymmetries
measured in $D^0\to K_S K_S$, as described in
Ref.~\cite{Grossman:2011zk}.  Adopting the convention $CP \ket{D^0} =
-\ket{\overline{D}{}^0}$ \cite{Gersabeck:2011xj} the amplitude of $\ov
D{}^0\to K_S K_S$ is
\begin{align}
\overline{\mathcal{A}} &= -\lambda_{sd}^* \mathcal{A}_{sd} +
\frac{\lambda_b^*}{2} \mathcal{A}_b. \label{eq:asdbbar} 
\end{align}
The direct CP asymmetry reads    
\begin{align}
a_{CP}^{\mathrm{dir}} &\equiv \frac{
        \vert \mathcal{A}\vert^2 - \vert \overline{\mathcal{A}} \vert^2
       }{
        \vert \mathcal{A}\vert^2 + \vert \overline{\mathcal{A}} \vert^2
         } \label{eq:CPasym1}\\
&=  \frac{ \mathrm{Im}\,\lambda_b}{\vert
  \mathcal{A}\vert }  
   \; 
  \mathrm{Im}\frac{
  \mathcal{A}_b}{\mathcal{A}_{sd}} \vert \mathcal{A}_{sd} \vert \, . 
  \label{eq:CPasym} 
\end{align}
Here and in the following we neglect terms of order $\lambda_b^2$ and
higher. Furthermore we use the PDG convention for the CKM elements, so that
$\lambda_{sd}$ is real and positive up to corrections of order
$\lambda_b$.

For the discussion of mixing-induced CPV we also follow the 
conventions of Ref.~\cite{Gersabeck:2011xj}: with the mass eigenstates
$\ket{D_{1,2}} = p \ket{D^0} \pm q\ket{\overline{D}^0}$ 
we define the weak phase $\phi$ governing CPV in the interference between the
\ddm\ and the $D^0\to K_S K_S$ decay through
\begin{align}
\frac qp \frac{\ov{\mathcal{A}}}{\mathcal{A}} &= 
-\frac{q}{p}\, \frac{\lambda_{sd}^*}{\lambda_{sd}} \, \frac{\ds 1 -
  \frac{\lambda_b^*}{2 \lambda_{sd}^*}
  \frac{\mathcal{A}_b}{\mathcal{A}_{sd}} }{\ds 1 - \frac{\lambda_b}{2
    \lambda_{sd}} \frac{\mathcal{A}_b}{\mathcal{A}_{sd}} }
\nn \\[1mm]
&\equiv\left| \frac qp \right| \,
 \left| \frac{\ov{\mathcal{A}}}{\mathcal{A}}\right| \, e^{i\phi} \,.
\label{eq:phi}
\end{align}
In this paper we focus on CPV effects which are specific to the decay
$D^0\to K_S K_S$. It is therefore useful to define a CP phase
$\phi_{\mathrm{mix}}$ which enters all mixing-induced CP asymmetries in
  a universal way: 
\begin{align}
-\frac{q}{p}\, \frac{\lambda_{sd}^*}{\lambda_{sd}}
&\equiv
\left|\frac{q}{p}\right| e^{i \phi_{\mathrm{mix}}}\,. 
\label{eq:phimix}
\end{align}
Comparing \eqsand{eq:phi}{eq:phimix} one verifies that
$\phi_{\mathrm{mix}}$ coincides with $\phi$ if one sets
$\lambda_b$ to zero in $\ov{\mathcal{A}}/\mathcal A$.
In the hunt for new physics (NP) in \ddm, which may well enhance 
$\phi_{\mathrm{mix}}$ over the SM expectation 
$\phi_{\mathrm{mix}}={\cal O} ({\imag} \lambda_b/\lambda_{sd})$,
one fits the CPV data of all available $D^0$ decays to a common
phase $\phi_{\mathrm{mix}}$ \cite{Grossman:2006jg,Amhis:2012bh}. 
In the case of $D^0\to K_SK_S$, however, we face the possibility 
that already the SM contributions lead to the situation 
$|\phi| \gg |\phi_{\mathrm{mix}}|$. Comparing \eq{eq:phi} with 
\eq{eq:phimix} one finds 
\begin{align}
\frac{\ds 1 -
  \frac{\lambda_b^*}{2 \lambda_{sd}^*}
  \frac{\mathcal{A}_b}{\mathcal{A}_{sd}} }{\ds 1 - \frac{\lambda_b}{2
    \lambda_{sd}} \frac{\mathcal{A}_b}{\mathcal{A}_{sd}} }
&= 
\left| \frac{\ov{\mathcal A}}{\mathcal A} \right| \, 
e^{i(\phi-\phi_{\mathrm{mix}})} \nn\\
&= \left(1- a_{CP}^{\mathrm{dir}} \right)\,
   e^{i(\phi-\phi_{\mathrm{mix}})}\,, \label{eq:comp}
\end{align}
where we have used \eq{eq:CPasym1}, discarding 
higher-order terms $\sim (a_{CP}^{\mathrm{dir}})^2$ as usual.
By expanding \eq{eq:comp} to first order in $
\lambda_b$ and $\phi-\phi_{\mathrm{mix}}$ we arrive at
\begin{align}
\phi-\phi_{\mathrm{mix}} &= 
  \imag \frac{\lambda_b}{\lambda_{sd}} \real 
 \frac{\mathcal A_b}{\mathcal A_{sd}} =
\frac{ \mathrm{Im}\,\lambda_b}{\vert
  \mathcal{A}\vert }  
   \; 
  \mathrm{Re}\frac{
  \mathcal{A}_b}{\mathcal{A}_{sd}} \vert \mathcal{A}_{sd} \vert \,.
\label{eq:cpm2}
\end{align}
\eqsand{eq:CPasym}{eq:cpm2} form the basis of the analysis presented in
the following sections. In \eqsand{eq:CPasym}{eq:cpm2} $|\mathcal A|$ is trivially 
related to the well-measured branching ratio:
{
\begin{align}
|\mathcal A| &= \sqrt{\frac{\mathcal{B}(D\rightarrow K_S
    K_S)}{\mathcal{P}(D,K,K)}}\,, \nn\\
\mathcal{P}(D,K,K) &\equiv\tau \, \frac{1}{ 16 \pi m_D^2} 
\sqrt{m_D^2 - 4 m_{K^0}^2 } \,, \label{eq:phasespace}
\end{align}} 
The experimental value is $\mathcal{B}(D^0\rightarrow K_SK_S) = (0.17\pm0.04)\cdot 10^{-3}$ \cite{Beringer:1900zz}.
The nontrivial quantities entering the predictions of
$a_{CP}^{\mathrm{dir}}$ and $\phi-\phi_{\mathrm{mix}}$ are
$\mathcal{A}_b$ and the phase of $\mathcal{A}_{sd}$.

The time-dependent CP asymmetry reads
\begin{align}
A_{CP}(t) &= \frac{\Gamma(D^0(t) \to K_{S} K_S) - \Gamma(\ov D{}^0(t) \to K_{{S}} K_S)
	}{
		\Gamma(D^0(t) \to K_{{S}} K_S) + \Gamma(\ov D{}^0(t) \to K_{{S}} K_S)} \nn\\ 
&= a_{CP}^{\mathrm{dir}} - A_{\Gamma} \frac{t}{\tau} \,.
\label{eq:time}
\end{align}
Here $\tau$ is the $D^0$ lifetime and 
\begin{align}
A_{\Gamma} {=} \left[ \frac{1}{2} \left(\left| \frac{q}{p}\right|^2
    - 1\right) - a_{CP}^{\mathrm{dir}} \right] y \cos\phi - x \sin\phi\, .
\label{eq:aga}
\end{align}
\eq{eq:aga} contains the mass difference $\dm$ and the width
  difference $\dg$ between the mass eigenstates $D_1$ and $D_2$  through 
  $x=\tau \dm$ and $y=\tau \dg/2$. {In \eqsand{eq:time}{eq:aga} all 
quadratic (and higher) terms in tiny quantities are neglected.}  
In time-integrated measurements, LHCb measures the quantity
\cite{Gersabeck:2011xj, Aaij:2014gsa,Aaij:2015yda}
\begin{align}
A_{CP} &= a_{CP}^{\mathrm{dir}} - A_{\Gamma} \frac{\langle t\rangle}{\tau},
\end{align}
where $\langle t\rangle$ is the average decay time.  CLEO has
  measured~\cite{Bonvicini:2000qm}
\begin{align}
A_{CP}^{\mathrm{CLEO}} &= -0.23 \pm 0.19 . 
\end{align}
{Recently LHCb has reported the preliminary} result
\cite{TalkAlexander:2015}
\begin{align}
A_{CP}^{\mathrm{LHCb}} &= -0.029 \pm 0.052 \pm 0.022 \,.   
\end{align}

\section{Topological amplitudes \label{sec:topological}}

\begin{figure}[t]
\begin{center}
\subfigure[\label{su3limit-exchange}]{
	\includegraphics[width=0.31\textwidth]{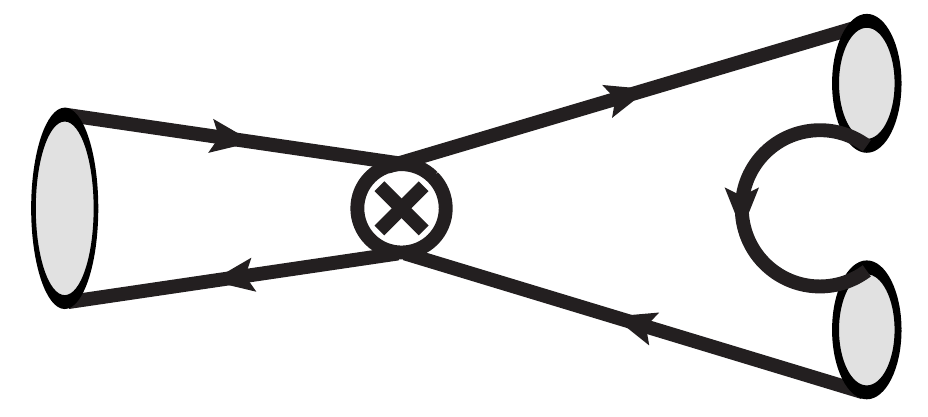}
	}
\subfigure[\label{su3limit-penguin-annihilation}]{
        \includegraphics[width=0.31\textwidth]{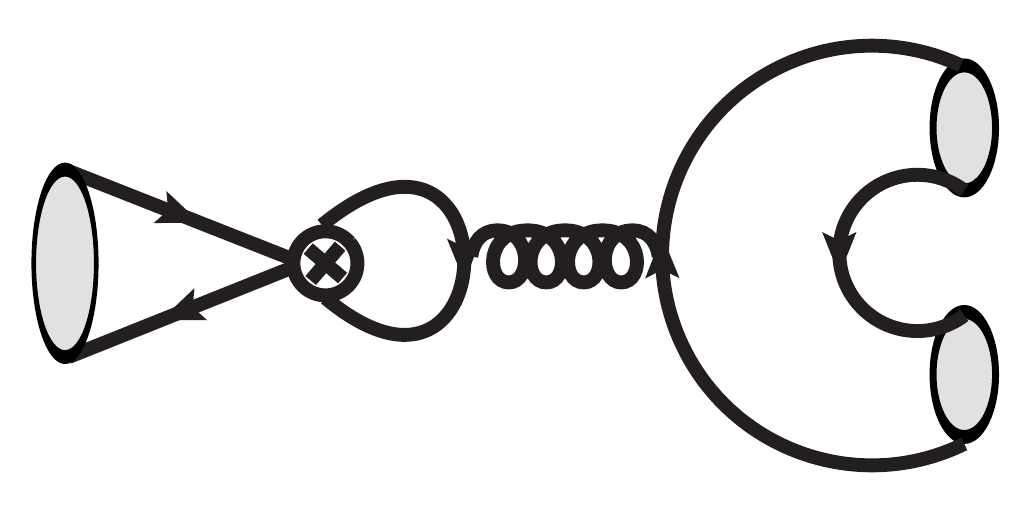}
	}
\end{center}
\caption{{Topological amplitudes:} 
(a) exchange ($E$) and (b) penguin annihilation ($PA$).
{Ref.~\cite{Atwood:2012ac} claims that 
$D^0\rightarrow K_SK_S$ is Zweig suppressed,
but this statement is only true for the PA diagram.}
\label{fig:su3limit-penguin-annihilation}
}
\end{figure}

The decomposition of $\mathcal{A}_{sd}$ and $\mathcal{A}_{b}$ in terms
of topological amplitudes reads \cite{Muller:2015lua}
\begin{align}
 \mathcal{A}_{sd} &= \frac{E_1+E_2-E_3}{\sqrt{2}}\,, \label{eq:decomp-1}\\
 \mathcal{A}_{b}  &= \frac{2 E + E_1 + E_2 + E_3 + PA}{\sqrt{2}} \label{eq:decomp0} \\
  &= -\mathcal{A}_{sd} + \frac{2 E_{KK} + PA}{\sqrt{2}}\,. \label{eq:decomp}
\end{align}
Here $E_{KK}\equiv E+E_1+E_2$ is the combination of exchange diagrams
appearing in $D^0\rightarrow K^+K^-$.  The exchange ($E$) and
penguin annihilation ($PA$) diagrams are shown in
Fig.~\ref{fig:su3limit-penguin-annihilation}.  {$E_{1,2,3}$ account
  for first-order SU(3)$_F$ breaking in diagrams containing $s$-quark
  lines (for their precise definition see Table~II of
  Ref.~\cite{Muller:2015lua}).}  As in Ref.~\cite{Muller:2015rna} $PA_q$
denotes the penguin annihilation diagram with quark $q$ running in the
loop. We use the combinations
\cite{Golden:1989qx,Pirtskhalava:2011va,Hiller:2012xm}
\begin{align}
\lambda_s& PA_s  + \lambda_d PA_d + \lambda_b PA_b 
= 
\nn\\
&\;\lambda_{sd} (PA_s-PA_d) 
 +  \frac{\lambda_s+\lambda_d}{2}\,\, (PA_s+PA_d- 2 PA_b) \\
&\equiv \lambda_{sd} PA_{\mathrm{break}} - \frac{\lambda_b}{2} PA\,
 \label{eq:PAconvention} .
\end{align}
{We recall that $E$, $E_{1,2,3}$, $PA$,\ldots are defined for 
$D^0\to K^0\ov K{}^0$ or $D^0\to K^+K^-$. Since 
$\mathcal A_{sd}$ and $\mathcal A_{b}$
instead involve $K_SK_S$, the factor of $-1/\sqrt2$ of \eq{eq:bose}
appears in \eqsto{eq:decomp-1}{eq:decomp}.}

Next we define the strong phase
\begin{align}
\delta \equiv \mathrm{arg}\left(\frac{2 E_{KK} + PA}{A_{sd}}\right) \,,
\end{align}
and the positive quantity
\begin{align}
  R 
&={- \frac{\mathrm{Im} \lambda_b}{|\mathcal A|}} 
 \frac{\vert 2 E_{KK} + PA
    \vert}{\sqrt{2}}. \label{eq:defineR}
\end{align}
{With \eq{eq:decomp} we can write \eq{eq:cpm2} as
\begin{align}
\phi - \phi_{\mathrm{mix}} &= 
   \imag \frac{\lambda_b}{\lambda_{sd}} 
   \, \real \frac{-\mathcal{A}_{sd}  + ( 2 E_{KK} + PA )/\sqrt{2}}{\mathcal{A}_{sd}}  
\nn \\
&= -\mathrm{Im}\frac{\lambda_b}{\lambda_{sd}} 
   - R \cos \delta  \,. \label{eq:indirectCP}
\end{align}
In the same way one finds} 
\begin{align}
a_{CP}^{\mathrm{dir}} &= {-} R
 \sin\delta\,. \label{eq:analyticestimate} 
\end{align}
{\eqsand{eq:indirectCP}{eq:analyticestimate} mean that 
$a_{CP}^{\mathrm{dir}}$ and $ \phi - \phi_{\mathrm{mix}}$ lie on a circle
  with radius $R$ centered at
  $(-\mathrm{Im}\frac{\lambda_b}{\lambda_{sd}},0)$. 
The allowed points are parametrized by the phase 
$\delta$, which we cannot predict. The actual value of $\delta$, however, 
is of minor importance for the discovery potential of CPV, because 
$\delta$ only controls how the amount of CPV is shared between 
$a_{CP}^{\mathrm{dir}}$ and $ \phi - \phi_{\mathrm{mix}}$. The crucial
  parameter is $R$, which determines the maximal values of 
$|a_{CP}^{\mathrm{dir}}|$ and  $|\phi - \phi_{\mathrm{mix}}|$.
Once  $a_{CP}^{\mathrm{dir}}$ and $ \phi -
    \phi_{\mathrm{mix}}$ are precisely measured one can determine $R$
through}
\begin{align}
R &= \sqrt{a_{CP}^{\mathrm{dir}}{}^2 + 
 \left( \phi - \phi_{\mathrm{mix}} + 
 \mathrm{Im}{\frac{\lambda_b}{\lambda_{sd}}} 
  \right)^2 } \,. \label{eq:circle} 
\end{align}
The experimental value can then be confronted with the theoretical
estimate presented in the next section. The impact of our estimate on
$a_{CP}^{\mathrm{dir}}$ and $\phi - \phi_{\mathrm{mix}}$ will be
presented below in Fig.~\ref{fig:circles}.

\section{Estimate of $\mathbf{\vert PA\vert}$ and
  $\mathbf{R}$ \label{sec:estimate}}

 The quantity $\vert E_{KK}\vert$ can be {determined from our global
   fit to} branching ratios \cite{Muller:2015lua}.  {For the
   calculation of $PA$ we exploit the large momentum $\sqrt{q^2}\sim
   1.5\gev$ flowing through the penguin loop in
   \fig{su3limit-penguin-annihilation} and calculate this loop
   perturbatively as in Ref.~\cite{Brod:2011re}. 
   Such methods are routinely used in $B$ physics \cite{Bander:1979px, Beneke:1999br, 
   Beneke:2001ev, Beneke:2003zv, Beneke:2004dp, Frings:2015eva}, but 
   their applicability to charm physics is not clear.} 

We work in a five-flavor theory, so that only current-current
  operators appear in the effective {Hamiltonian}. With 
$Q_2\equiv ({\ov{u} s})_{V-A} (\ov{{s}} c)_{V-A}+ ({\ov{u} d} )_{V-A} (\ov d
c)_{V-A} -2 ({ \ov u b})_{V-A} (\ov b c)_{V-A}$ {and the Wilson
coefficient $C_2\sim 1.2$} we may write 
\begin{align}
\mathcal{A}_b & = \frac{G_F}{\sqrt{2}} {C_2} \bra{K_SK_S} Q_2 \ket{D^0},
\label{eq:ham}
\end{align}
because the contribution of the color-flipped operator $Q_1$ is highly
suppressed. For our estimate of the ratio $PA/E_{KK}$ in this section we adopt  
the SU(3)$_F$ limit and identify $E_{KK}$ with $E$. In this limit we can
combine \eqsand{eq:ham}{eq:decomp0} into 
\begin{align}
\frac{G_F}{\sqrt{2}} {C_2} \bra{K_SK_S} Q_2 \ket{D^0} &= 
                       \frac{2 E +PA}{\sqrt{2}}\,.  \label{eq:hep}
\end{align}
The penguin diagram can be written as \cite{Lenz:1997aa}
\begin{align}
  & PA =
  {G_F} \frac{\alpha_s}{4\pi} C_2 \times \nn\\
  &\quad \times \sum_{i=3}^6 \left( r_{2i}^d  +
    r_{2i}^s - 2 r_{2i}^b \right)  
    \bra{K_SK_S} Q_i \ket{D^0} \,, \label{eq:factorPingu}
\end{align}
with the loop function $r_{24}^q \equiv r_{24}^q(q^2, m_q^2, \mu^2 ) =
r_{26}^q$ defined in Ref.~\cite{Lenz:1997aa}. {$\mu\sim \sqrt{q^2}$ is the renormalization 
scale which also enters $\alpha_s$ and $C_2$ in \eq{eq:factorPingu}.
$Q_{3-6}$ are the usual four-quark penguin operators, we will need
\begin{align}
Q_{4,6}&= (\ov u{}^\alpha c^\beta)_{V-A} \!\!\!\sum_{q=u,d,s,c{,b}} 
  (\ov q{}^\beta
  q^\alpha)_{V\mp A}. \label{eq:q46}
\end{align}
$PA$ is
  color-suppressed w.r.t.\ $E$ and this suppression is encoded in
  \eq{eq:factorPingu} through $\alpha_s\sim 1/N_c$. The contributions
  from the matrix elements $ \langle Q_{3,5} \rangle$ are further
  suppressed and are neglected in the following. We write}  
$\langle Q_4\rangle + \langle Q_6 \rangle = -2
\left( M_{VA}^{{d}} +  M_{VA}^{{s}} \right)$ with
\begin{align}
M_{VA}^{{q}} &\equiv \bra{K_{{S}} K_{{S}}} (\bar{q}_{\alpha} q_{\beta} )_V (
\bar{u}_{\beta} c_{\alpha} )_A \ket{D^0}\,. \label{eq:hadronicME}
\end{align}
The other quark flavors in the sum in \eq{eq:q46} contribute to 
$D^0\to K_{{S}}K_{{S}}$ only through another loop diagram, yielding a
contribution of higher order in {$\alpha_s$}.  
With 
\begin{align}
p &\equiv r_{24}^d + r_{24}^s - 2 r_{24}^b\,, 
\end{align}
we can write $PA$ in a compact form:
\begin{align}
PA = -{G_F} \frac{\alpha_s}{\pi} C_2 \, p\, M_{VA}^d   \,, 
\label{eq:pa}
\end{align}
where we have invoked the SU(3)$_F$ limit to set $M_{VA}^d=M_{VA}^s$.
The $\mu$-dependence cancels in $p$, which furthermore does not depend
on $m_c$ in the considered leading order. It is an excellent numerical
approximation to expand $p$ to first order in $m_s^2/q^2$ and
$q^2/m_b^2$ (while setting $m_d=0$). The expanded expression reads
\begin{align} 
 p&= -\frac{10}{9} - \frac23 i \pi  - \frac{2 m_s^2}{q^2} + \frac{2
   q^2}{15 m_b^2} + \frac23 \ln \frac{q^2}{m_b^2} {\,.} 
\label{eq:p}
\end{align} 
  It is worthwhile to discuss how this result translates into an
  expression in a four-flavor theory, in which the $b$ quark is
  integrated out at the scale $\mu_b={\cal O}(m_b)$: in this alternative
  approach the piece $-2r_{2i}^b$ of \eq{eq:factorPingu} resides in the
  initial conditions of the penguin coefficients $C_{3-6}$ generated at
  $\mu_b$. The four-flavor theory permits the use of the
  renormalization group (RG) to resum the log $\ln(\mu_b/\sqrt{q^2})$ to
  all orders in perturbation theory, but this resummation is inconsistent
  since $\ln(\mu_b/\sqrt{q^2})$ is smaller than the nonlogarithmic
  terms in $-2r_{2i}^b$. Without RG summation the four-flavor theory
  reproduces exactly the analytic result in \eq{eq:p}, which is
  independent of renormalization scale and scheme.

To estimate $M_{{VA}}^{{d}}$ we want to relate it to $E$ using \eq{eq:hep}. 
After Fierz-rearranging $Q_2$ we can express {the} LHS of \eq{eq:hep} 
in terms of $M_{VA}^q$ and
\begin{align} 
  M_{AV}^{{q}} &\equiv \bra{K_SK_S} (\bar{q}_{\alpha} q_{\beta} )_A (
  \bar{u}_{\beta} c_{\alpha} )_V \ket{D^0}\, .  \label{eq:hadronicME2}
\end{align}
The exchange topology {reads (cf.\ \eq{eq:hep})} 
\begin{align} 
 E&= {G_F C_2}\bra{K_SK_S} ({\ov u d})_{V-A} (\ov d c)_{V-A}  \ket{D^0} - PA_d  \nn \\
 &= {-G_F C_2} \left( M_{AV}^d+M_{VA}^d\right) - {G_F\frac{\alpha_s}{\pi}} C_2  r_{24}^d M^d_{VA} \,.
\label{eq:emva}
\end{align}
To leading order in $\alpha_s$ {we have} therefore $E={-G_F C_2}\left(M_{AV}^d+M_{VA}^d\right)$. For the
desired estimate of $PA/E$ we need $M_{VA}^d/E$. We can place a bound on 
this quantity with \eq{eq:emva}, if we assume that $ |M^d_{VA}|$  
is not much larger than $| M_{AV}^d+M_{VA}^d |$; i.e.\ we do not
consider the case of large cancellations between $M_{AV}^d$ and
$M_{VA}^d$ in $E$. In view of the fact that $E$ is numerically large
\cite{Muller:2015lua} this assumption seems justified. Writing 
\begin{align} 
 M^d_{VA} &= \kappa (M_{AV}^d+M_{VA}^d) {\,,}
\label{eq:kap}
\end{align}
we vary $|\kappa|$ between 0 and 2. Now \eq{eq:emva} entails 
\begin{align} 
\frac{M^d_{VA}}{E} &= { \frac{\kappa}{-G_F C_2 (1 + \kappa \frac{\alpha_s}{\pi} r_{24}^d)}} {\,,}
\label{eq:mest}
\end{align}
and thus 
\begin{align}
\left|2 E_{KK} + PA\right| &= \left| 2 E_{KK}\right| 
	\left| 
	1 + \frac{\alpha_s}{2\pi} p \frac{\kappa}{1 + \kappa \frac{\alpha_s}{\pi} r_{24}^d}
	\right| \\
	&\leq   \left| 2 E_{KK}\right| \times 1.3\,, \label{eq:PAestimate}
\end{align}
{Here} we {have used} $\mu=\sqrt{q^2} = 1.5$~GeV, $m_s{(\mu)} =
0.104$~GeV, $m_b{(\mu)} = 4.18$~GeV, and $\alpha_s{(\mu)} =
0.328$. {($r_{24}^d$ is evaluated in the NDR scheme.)} 
Inserting {finally} Eq.~(\ref{eq:PAestimate}) into
Eq.~(\ref{eq:defineR}) gives the upper limit
\begin{align}
  R\, &{\leq}  - {1.3} \frac{\mathrm{Im} \lambda_b}{|\mathcal A|} 
 \frac{\vert 2 E_{KK} \vert}{\sqrt{2}} \, . 
 \label{eq:radiusbound}
\end{align}
{This bound determines the radius of the circle which defines the
  allowed area for $(\phi - \phi_{\mathrm{mix}},a_{CP}^{\mathrm{dir}})$
  via Eq.~(\ref{eq:circle}). I.e.\ \eq{eq:radiusbound} determines the
  maximal size of both $|a_{CP}^{\mathrm{dir}}|$ and $|\phi -
  \phi_{\mathrm{mix}}|$ (neglecting the small
  $\imag\lambda_b/\lambda_{sd}$ in \eq{eq:indirectCP}).} {If future
  data violate Eq.~(\ref{eq:radiusbound}), this will signal new physics or a
  dynamical enhancement of $PA$ over the perturbative result in
  \eq{eq:pa}. Sec.~\ref{sec:pheno} discusses how these two scenarios can
  be distinguished with the help of other measurements.}

The relation of $r_{24}^q(q^2, m_q^2, \mu^2)$ to $G(s,x)$ in
Ref.~\cite{Beneke:2001ev} is given as
\begin{align}
r_{24}^q(q^2, m_q^2, \mu^2 ) &=  \frac{1}{3} - 
   \frac{1}{3} \log\left(\frac{\mu^2}{m^2}\right) \nn \\ 
&\quad	
{ - }
\frac{1}{2} G\left(\frac{m_q^2 - i \varepsilon}{m^2}, 
\frac{q^2}{m^2}\right) \,, \label{eq:r24andG}
\end{align}
with an arbitrary mass $m^2$. Note that the $-i\varepsilon$ prescription
is essential here; an erroneous omission of this small imaginary part 
results in a numerically large mistake.   
The prefactor of $G(x,y)$ in Eq.~(\ref{eq:r24andG}) disagrees with 
Ref.~\cite{Brod:2011re}. {We further find that the $b$-quark contribution 
$-2 r_{24}^b$ is numerically as important as $r_{24}^d+r_{24}^s$:}
\begin{align}
 r_{24}^d(q^2, 0, \mu^2 )   &= -0.22 - i\, 1.05 \,,\\
 r_{24}^s(q^2, m_s^2, \mu^2 )   &= -0.23 - i\, 1.05 \,,\\
-2 r_{24}^b(q^2, m_b^2, \mu^2 ) &= -2.02  \,.
\end{align} 

\section{Phenomenology \label{sec:pheno}} 
{The last element needed for the calculation of our bound in
  \eq{eq:radiusbound} is $|E_{KK}|$. To find $|E_{KK}|$ we employ our
  global fit to all available branching ratios of $D$ decays to two
  pseudoscalar mesons \cite{Muller:2015lua}.}  Note that the {main
  constraint on this quantity stems from} $\mathcal{B}(D^0\rightarrow
K^+K^-)$ {(see} Table~III of Ref.~\cite{Muller:2015lua}). {The $D$
  decays entering our fit involve other topological amplitudes in
  addition to $E$ and $PA$; in the following we refer to the
  color-favored tree (T), color-suppressed tree (C), annihilation (A) 
  and penguin (P) amplitudes. } 

We {consider} two scenarios: in the first {scenario} the
SU(3)$_F$-limit amplitudes $C$ and $E$ are varied completely
free{ly}. In the second {scenario} we {apply} $1/N_c$ counting
\cite{'tHooft:1973jz,Buras:1985xv,Buras:1998ra} to the amplitudes,
{where $N_c=3$ is the number of colors. To leading order in $1/N_c$
  one can factorize $T$ which results in}
\begin{align}
  T^{\mathrm{fac}} &\equiv \frac{G_F}{\sqrt{2}} a_1 f_{\pi}
  \left(m_D^2-m^2_{\pi}\right) F_0^{D\pi}(m_{\pi}^2)\,.
\end{align}
\begin{figure*}[t]
\begin{center}
\subfigure[\label{fig:plot-1d-64}]{
        \includegraphics[width=0.47\textwidth]{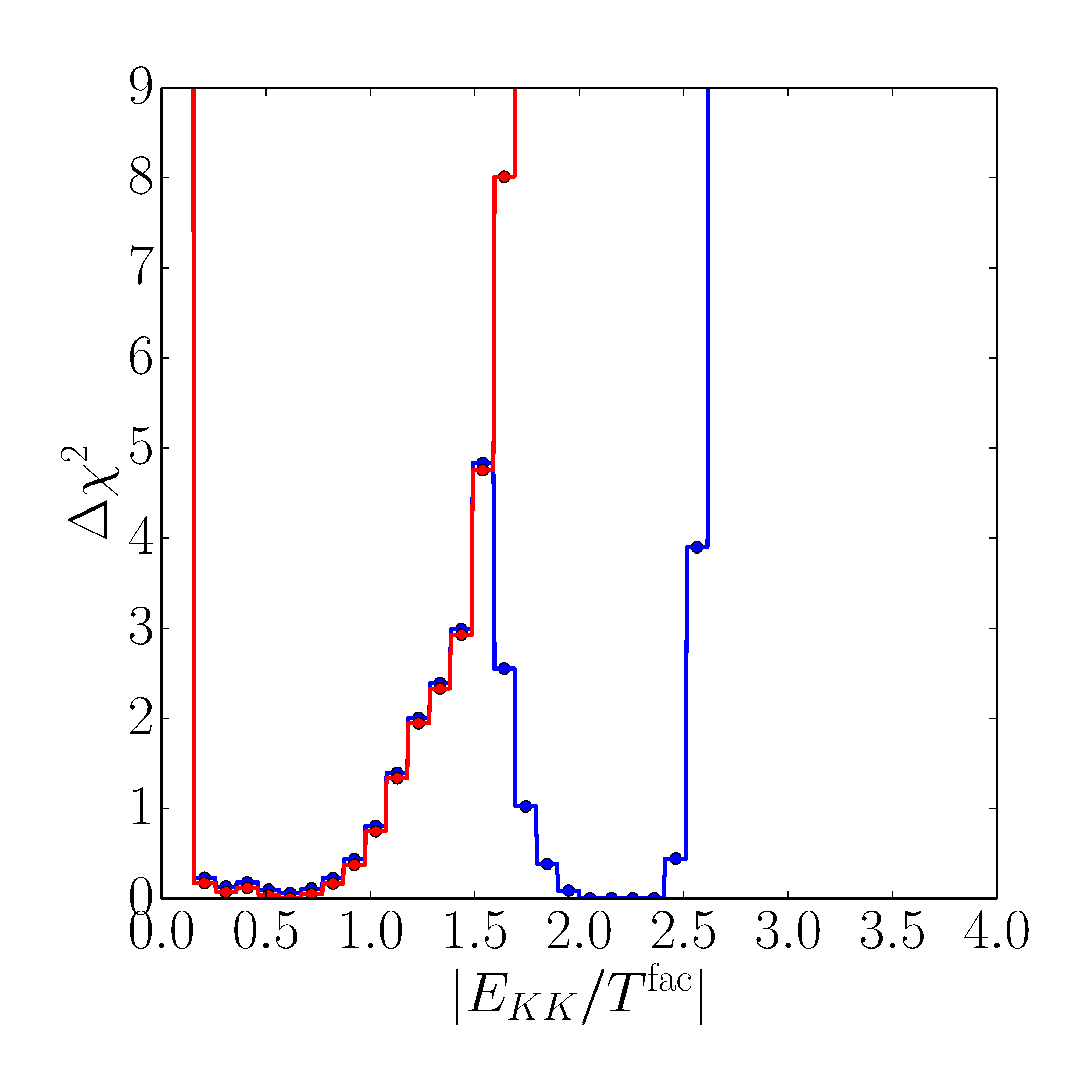}
	}
\subfigure[\label{fig:plot-1d-65}]{
        \includegraphics[width=0.47\textwidth]{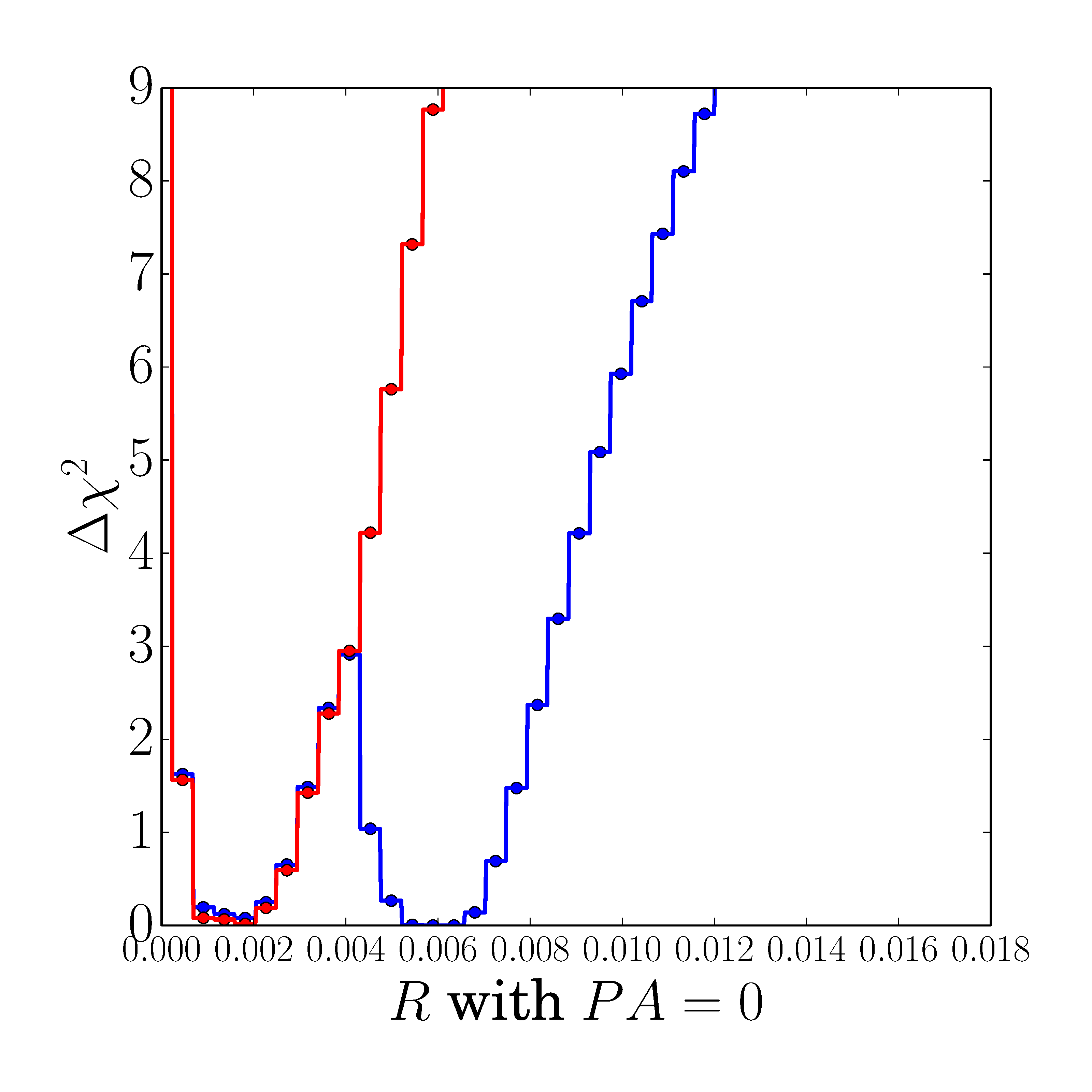}
	}
\end{center}
\caption{{(a)} $\Delta\chi^2$ profile of  $\vert E_{KK}\vert$. 
  {(b)}  $\Delta\chi^2$ profile of $
  R$ (defined in \eq{eq:defineR}) {for} $PA=0$.  
  {The blue and red curves correspond to the scenarios without and
    with $1/N_c$ counting applied to} 
  $C$ and $E$. Note that the red line lies partially
  on top of the blue line. \label{fig:deltachi2} }
\end{figure*}
{Here $a_1=1.06$ is the appropriate combination of Wilson
  coefficients, $m_\pi$ and $f_{\pi}$ are the mass and the decay constant of the
  pion, respectively, and $ F_0^{D\pi}$ is the appropriate $D\to \pi$
  form factor. (Recall that the SU(3)$_F$-limit amplitudes are defined
  for decays into pions.) In our second scenario} we assume that $\vert
(C+\delta_A)/T^{\mathrm{fac}}\vert,\, \vert
(E+\delta_A)/T^{\mathrm{fac}}\vert\leq 1.3$~\cite{Muller:2015rna}, where
$\delta_A$ parametrizes $1/N_c^2$ corrections {to the factorized
  annihilation (A) topology} \cite{Muller:2015lua}.

\begin{figure}[t]
\begin{center}
        \includegraphics[width=0.49\textwidth]{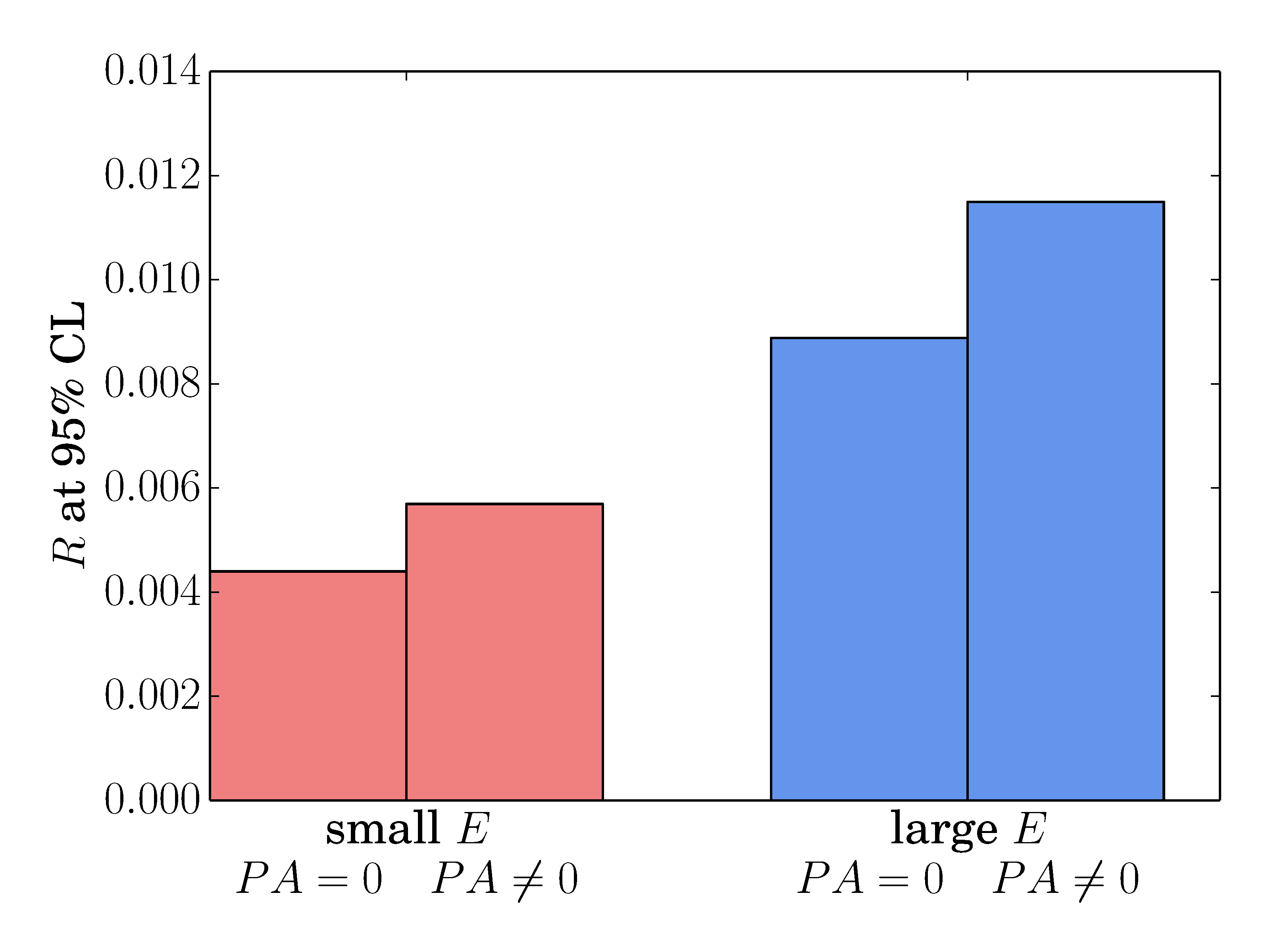}
\end{center}
\caption{Theoretical upper bounds on $R$. 
Predictions with (without) $1/N_c$ {counting
are labeled} \lq\lq{}small $E$\rq\rq{} (\lq\lq{}large $E$\rq\rq{}). To
{visualize} the contribution from exchange diagrams, we {also show} 
the result {for} $PA=0$. 
The case $PA\neq 0$ {is based on} the estimate in \eq{eq:radiusbound}.
\label{fig:barchart}
}
\end{figure}

The $\Delta \chi^2$-profile of $\vert E_{KK}/T^{\mathrm{fac}}\vert$
{returned by our global fit is shown in
  Fig.~\ref{fig:deltachi2}(a). Fig.~\ref{fig:deltachi2}(b) shows the
  $\Delta \chi^2$-profile of $R$ for the special case $PA=0$, in which
  the whole effect comes from the exchange diagram $E_{KK}$. }  The
corresponding 95\% C.L. bounds on $\vert E_{KK}/T^{\mathrm{fac}}\vert$
and $R$ inferred from Fig.~\ref{fig:deltachi2} and \eq{eq:radiusbound}
are given in Table~\ref{tab:results} and illustrated in
Fig.~\ref{fig:barchart}.  Note that {we do not treat}
$T^{\mathrm{fac}}$ {as} constant, but {also fit the form factor $
  F_0^{D\pi}$. Likewise our fit permits} $\mathcal{B}(D^0\rightarrow
K_SK_S)$ {to float within the experimental errors}. 
\begin{table}[t]
\begin{center}
\begin{tabular}{rcccc}
  \hline \hline
  &\multicolumn{2}{c}{with $1/N_c$}  &  \multicolumn{2}{c}{without $1/N_c$} \\
  & $PA=0$  & $PA\neq 0$~~ &  $PA= 0$  & $PA\neq 0$ \\ 
  $\vert E_{KK}/T^{\mathrm{fac}}\vert \;\;{\leq}$~~&  
  \multicolumn{2}{c}{1.5} & \multicolumn{2}{c}{2.6}  \\

$R\;\;{\leq}$~~& {0.004} & {0.006} & {0.009}  & {0.011}    \\\hline\hline
\end{tabular}
\caption{{$95\%$ C.L. upper} limits 
 ($\Delta\chi^2=3.84$), with or without $1/N_c$ input for $C$ and $E$. 
  \label{tab:results}}
\end{center}
\end{table}

\begin{figure*}[t]
\begin{center}
        \includegraphics[width=0.8\textwidth]{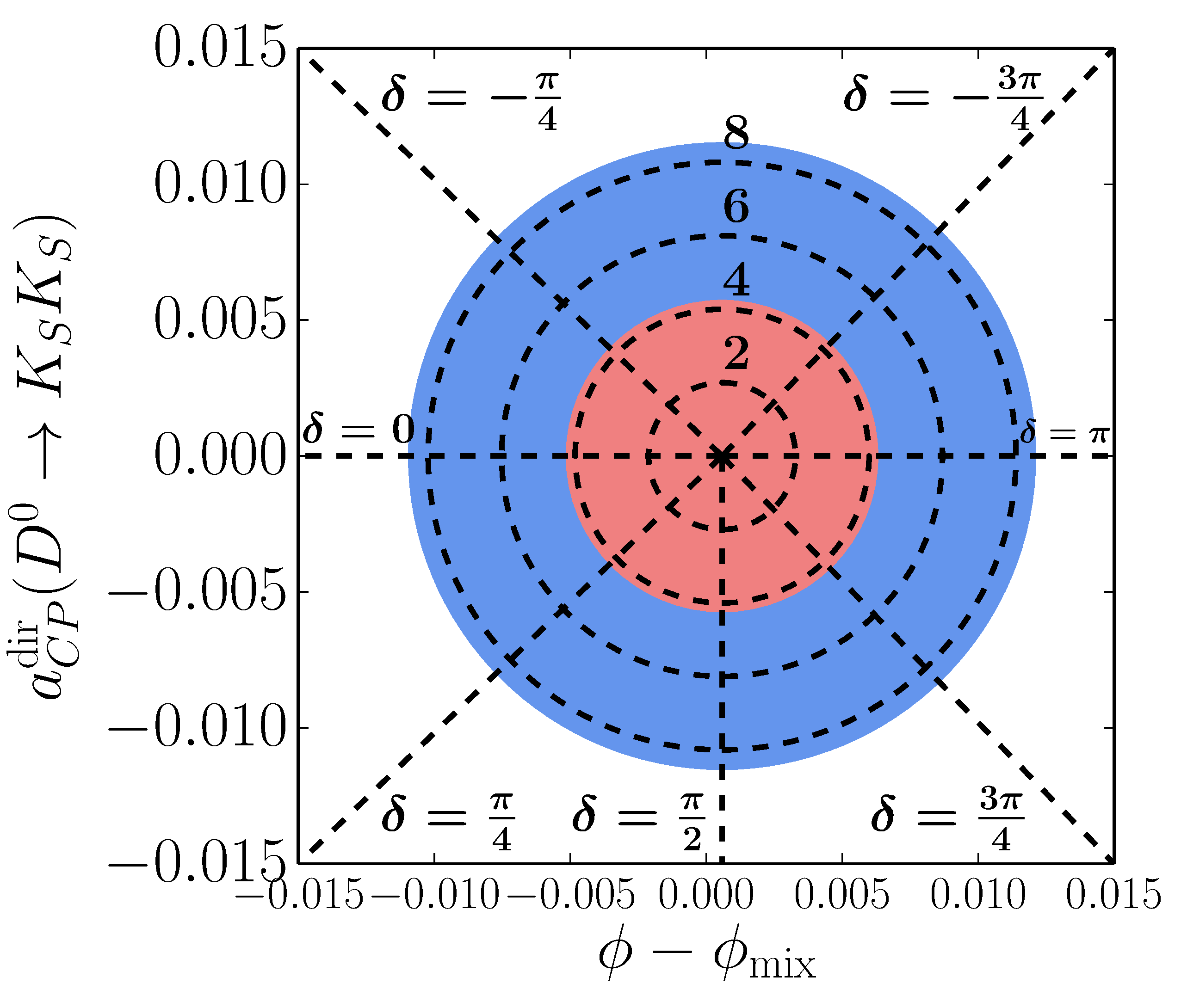}
\end{center}
\caption{
  {Correlation of} direct and {mixing-induced} CPV for
  $D^0\rightarrow K_SK_S$ from {\eqsto{eq:indirectCP}{eq:circle}.
    The one-dimensional} $95\%$ C.L. ($\Delta\chi^2=3.84$) upper limits on
  $a_{CP}^{\mathrm{dir}}(D^0\rightarrow K_SK_S)$ and
  $\phi-\phi_{\mathrm{mix}}$ are {shown in blue. If in addition
    $1/N_c$ counting is applied to the topological amplitudes $C$ and
    $E$, the allowed region shrinks to the red area.}  The black dashed
  lines show the radii which are obtained when setting $|2 E_{KK}
  +PA|/{2.52\cdot 10^{-6}\gev}$ to the annotated values. {The
    chosen reference value is the typical size of a factorized tree
    amplitude in $D$ decays, $T^{\mathrm{fac}} =2.52\cdot 10^{-6}$~GeV.
Further, for the black dashed lines  
$\mathcal{B}(D^0\rightarrow K_SK_S) = 0.17\cdot 10^{-3}$ is used.  The}
circle {is centered at}
$(-\mathrm{Im}(\lambda_b)/\lambda_{sd},0) = ( 6\cdot 10^{-4},0)$.
\label{fig:circles}
}
\end{figure*}
{\fig{fig:circles} condenses the main results of this paper into a
  single plot: the radial lines correspond to fixed values of the strong
  phase $\delta$ in \eqsand{eq:indirectCP}{eq:analyticestimate} in the
  $\phi-\phi_{\mathrm{mix}}$--$a_{CP}^{\mathrm{dir}}$ plane. The red and
  blue discs show the allowed regions for the two considered scenarios.}
 Note that our bounds depend on branching ratio measurements only and
 {do not involve} correlations to other CP asymmetries.  { The black
  circles correspond to different values of $\vert 2 E_{KK} + PA \vert$
  in \eq{eq:defineR}.  Future data on $\phi-\phi_{\mathrm{mix}}$ and
  $a_{CP}^{\mathrm{dir}}$ will allow us to determine $\delta$ and
  $R$. The experimental value of $R$ can then be confronted with the
  upper limits in Table~\ref{tab:results} to probe the color counting in
  $E_{KK}$ and our estimate of $PA$. New physics will mimic a dynamical
  enhancement of $PA$. In case an anomalously large value of $R$ will be
  found, one can proceed in the following way to discriminate between different
   explanations:
\begin{itemize}
\item[(i)] Several CP asymmetries involve $PA$, but do not grow with $|E|$. For
  example, 
\begin{align}
& a_{CP}^{\mathrm{dir}}(D^0\rightarrow K^+K^-),\, a_{CP}^{\mathrm{dir}}(D^0\rightarrow \pi^+\pi^-),\nn\\ 
& a_{CP}^{\mathrm{dir}}(D^0\rightarrow \pi^0\pi^0)\,, \label{eq:modes1}
\end{align}
all depend on $P+PA$ and are expected to be enhanced with $PA$ as well, unless the
increase is compensated by $-P$. But in this case
instead 
\begin{align}
& a_{CP}^{\mathrm{dir}}(D^+\rightarrow K_SK^+), a_{CP}^{\mathrm{dir}}(D_s^+\rightarrow K_S \pi^+),\nn\\
& a_{CP}^{\mathrm{dir}}(D_s^+\rightarrow K^+\pi^0)\,, \label{eq:modes2}
\end{align}
which involve $P$ rather than $P+PA$, become large. Thus a breakdown of
color counting in $E_{KK}$ can be distinguished from an enhanced $PA$.    
\item[(ii)] $PA$ can be enhanced by QCD dynamics or by new physics. In
  the first case the CP asymmetries in \eqsand{eq:modes1}{eq:modes2}   
  will still obey the sum rules of Ref.~\cite{Muller:2015rna}. 
  New physics will violate these sum rules if it couples differently 
  to down and strange quarks. 
\end{itemize}
}

{We close this section by comparing our result with other estimates 
of  $a_{CP}^{\mathrm{dir}}(D^0\rightarrow K_SK_S)$ in the literature. 
Using generic  SU(3)$_F$ counting Ref.~\cite{Brod:2011re} quotes}
\begin{align}
  \vert a_{CP}^{\mathrm{dir}}(D^0\rightarrow K_SK_S)\vert \lesssim \frac{2 \vert V_{cb}
    V_{ub}\vert }{\varepsilon \vert V_{cs} V_{us} \vert} \sim 0.6\%\,,
\end{align}
{where $\varepsilon$ quantifies} SU(3)$_F$ breaking. {Our result
  in Table~\ref{tab:results} agrees with this estimate. However,
  if the possibility of a large,  $1/N_c$-unsuppressed $|E_{KK}|$ is 
realized in nature, $|a_{CP}^{\mathrm{dir}}|$ can be twice as large.}  

{Ref.~\cite{Hiller:2012xm} relates $a_{CP}^{\mathrm{dir}}(D^0\rightarrow K_SK_S)$ to $\Delta
  a_{CP}^{\mathrm{dir}}\equiv a_{CP}^{\mathrm{dir}}(K^+K^-)- a_{CP}^{\mathrm{dir}}(\pi^+\pi^-)$.  
With present data this relation reads}
\begin{align}
  \vert a_{CP}^{\mathrm{dir}}\vert \lesssim \frac{3}{2} \times \Delta
  a_{CP}^{\mathrm{dir}} = 0.4\%\,. \label{eq:previous}
\end{align}
{This estimate assumes that two matrix elements corresponding to
  different SU(3)$_F$ representations are similar in magnitude. We
  remark that there is no} strict correlation between
$a_{CP}^{\mathrm{dir}}(D^0\rightarrow K_SK_S)$ and $\Delta
a_{CP}^{\mathrm{dir}}$, {because the two quantities involve different
  topological amplitudes.}

\section{Conclusions}
{We have studied the direct and
  mixing-induced CP asymmetries in $D^0\to K_S K_S$ in the Standard
  Model.  The allowed region for the corresponding two quantities
  $a_{CP}^{\mathrm{dir}}$ and $\phi-\phi^{\mathrm{mix}}$ is a disc whose 
  radius can be calculated in terms of the exchange amplitude $E_{KK}$ 
  and the penguin annihilation amplitude $PA$. We estimate $PA/E_{KK}$ 
  with a perturbative calculation and obtain $E_{KK}$ from a global fit 
  to $D$ branching fractions as described in Ref.~\cite{Muller:2015lua}.}
We find 
\begin{align}
\vert a_{CP}^{\mathrm{dir}}\vert 			   &\leq 1.1\%\quad (95\%\, \mathrm{C.L.})\,,\\
\vert \phi - \phi_{\mathrm{mix}} + \mathrm{Im}\frac{\lambda_b}{\lambda_{sd}} \vert &\leq 1.1\%\quad (95\%\, \mathrm{C.L.})\,.
\end{align}
{A simultaneous measurement of $ a_{CP}^{\mathrm{dir}}$ and $\phi -
  \phi_{\mathrm{mix}}$ will determine $|2E_{KK}+PA|$. A violation of the
bound $$\sqrt{a_{CP}^{\mathrm{dir}}{}^2 + 
 \left( \phi - \phi_{\mathrm{mix}} + 
 \mathrm{Im}{\frac{\lambda_b}{\lambda_{sd}}} 
  \right)^2 }\leq 1.1\%$$ will point to an anomalously enhanced $PA$.}  
In {this case} other CP asymmetries will also be enhanced.

\textbf{Note added in Proof:} 
The authors of Ref.~\cite{Brod:2011re} have informed us that they agree with our expression Eq.~(\ref{eq:r24andG}). 
The apparent difference is due to a typo in Eq.~(13) of Ref.~\cite{Brod:2011re}.
By comparing our numerical codes we could trace our numerical differences back to the 
\lq\lq{}$i \varepsilon$ problem\rq\rq{} mentioned at the end of Sec.~\ref{sec:estimate}.

\begin{acknowledgments}
  We thank Philipp Frings and Tim Gershon for useful discussions and the authors
of Ref.~\cite{Brod:2011re} for a thorough comparison of the penguin loop
function. UN and StS acknowledge the kind hospitality of the \emph{Munich Institute
    for Astro- and Particle Physics}.  The fits are performed using the
  \texttt{python} version of the software package
  \texttt{myFitter}~\cite{Wiebusch:2012en}. The Feynman diagrams are
  drawn using \texttt{Jaxodraw}~\cite{Binosi:2003yf,Vermaseren:1994je}.
  The presented work is supported by BMBF under contract no.~05H15VKKB1.
\end{acknowledgments}

\bibliography{unsts-15.bib}

\begin{thebibliography}{42}%
\makeatletter
\providecommand \@ifxundefined [1]{%
 \@ifx{#1\undefined}
}%
\providecommand \@ifnum [1]{%
 \ifnum #1\expandafter \@firstoftwo
 \else \expandafter \@secondoftwo
 \fi
}%
\providecommand \@ifx [1]{%
 \ifx #1\expandafter \@firstoftwo
 \else \expandafter \@secondoftwo
 \fi
}%
\providecommand \natexlab [1]{#1}%
\providecommand \enquote  [1]{``#1''}%
\providecommand \bibnamefont  [1]{#1}%
\providecommand \bibfnamefont [1]{#1}%
\providecommand \citenamefont [1]{#1}%
\providecommand \href@noop [0]{\@secondoftwo}%
\providecommand \href [0]{\begingroup \@sanitize@url \@href}%
\providecommand \@href[1]{\@@startlink{#1}\@@href}%
\providecommand \@@href[1]{\endgroup#1\@@endlink}%
\providecommand \@sanitize@url [0]{\catcode `\\12\catcode `\$12\catcode
  `\&12\catcode `\#12\catcode `\^12\catcode `\_12\catcode `\%12\relax}%
\providecommand \@@startlink[1]{}%
\providecommand \@@endlink[0]{}%
\providecommand \url  [0]{\begingroup\@sanitize@url \@url }%
\providecommand \@url [1]{\endgroup\@href {#1}{\urlprefix }}%
\providecommand \urlprefix  [0]{URL }%
\providecommand \Eprint [0]{\href }%
\providecommand \doibase [0]{http://dx.doi.org/}%
\providecommand \selectlanguage [0]{\@gobble}%
\providecommand \bibinfo  [0]{\@secondoftwo}%
\providecommand \bibfield  [0]{\@secondoftwo}%
\providecommand \translation [1]{[#1]}%
\providecommand \BibitemOpen [0]{}%
\providecommand \bibitemStop [0]{}%
\providecommand \bibitemNoStop [0]{.\EOS\space}%
\providecommand \EOS [0]{\spacefactor3000\relax}%
\providecommand \BibitemShut  [1]{\csname bibitem#1\endcsname}%
\let\auto@bib@innerbib\@empty
\bibitem [{\citenamefont {Fanti}\ \emph {et~al.}(1999)\citenamefont {Fanti}
  \emph {et~al.}}]{Fanti:1999nm}%
  \BibitemOpen
  \bibfield  {author} {\bibinfo {author} {\bibfnamefont {V.}~\bibnamefont
  {Fanti}} \emph {et~al.} (\bibinfo {collaboration} {NA48}),\ }\href {\doibase
  10.1016/S0370-2693(99)01030-8} {\bibfield  {journal} {\bibinfo  {journal}
  {Phys. Lett.}\ }\textbf {\bibinfo {volume} {B465}},\ \bibinfo {pages} {335}
  (\bibinfo {year} {1999})},\ \Eprint {http://arxiv.org/abs/hep-ex/9909022}
  {arXiv:hep-ex/9909022 [hep-ex]} \BibitemShut {NoStop}%
\bibitem [{\citenamefont {Lai}\ \emph {et~al.}(2001)\citenamefont {Lai} \emph
  {et~al.}}]{Lai:2001ki}%
  \BibitemOpen
  \bibfield  {author} {\bibinfo {author} {\bibfnamefont {A.}~\bibnamefont
  {Lai}} \emph {et~al.} (\bibinfo {collaboration} {NA48}),\ }\href {\doibase
  10.1007/s100520100822} {\bibfield  {journal} {\bibinfo  {journal} {Eur. Phys.
  J.}\ }\textbf {\bibinfo {volume} {C22}},\ \bibinfo {pages} {231} (\bibinfo
  {year} {2001})},\ \Eprint {http://arxiv.org/abs/hep-ex/0110019}
  {arXiv:hep-ex/0110019 [hep-ex]} \BibitemShut {NoStop}%
\bibitem [{\citenamefont {Batley}\ \emph {et~al.}(2002)\citenamefont {Batley}
  \emph {et~al.}}]{Batley:2002gn}%
  \BibitemOpen
  \bibfield  {author} {\bibinfo {author} {\bibfnamefont {J.~R.}\ \bibnamefont
  {Batley}} \emph {et~al.} (\bibinfo {collaboration} {NA48}),\ }\href {\doibase
  10.1016/S0370-2693(02)02476-0} {\bibfield  {journal} {\bibinfo  {journal}
  {Phys. Lett.}\ }\textbf {\bibinfo {volume} {B544}},\ \bibinfo {pages} {97}
  (\bibinfo {year} {2002})},\ \Eprint {http://arxiv.org/abs/hep-ex/0208009}
  {arXiv:hep-ex/0208009 [hep-ex]} \BibitemShut {NoStop}%
\bibitem [{\citenamefont {Alavi-Harati}\ \emph {et~al.}(1999)\citenamefont
  {Alavi-Harati} \emph {et~al.}}]{AlaviHarati:1999xp}%
  \BibitemOpen
  \bibfield  {author} {\bibinfo {author} {\bibfnamefont {A.}~\bibnamefont
  {Alavi-Harati}} \emph {et~al.} (\bibinfo {collaboration} {KTeV}),\ }\href
  {\doibase 10.1103/PhysRevLett.83.22} {\bibfield  {journal} {\bibinfo
  {journal} {Phys. Rev. Lett.}\ }\textbf {\bibinfo {volume} {83}},\ \bibinfo
  {pages} {22} (\bibinfo {year} {1999})},\ \Eprint
  {http://arxiv.org/abs/hep-ex/9905060} {arXiv:hep-ex/9905060 [hep-ex]}
  \BibitemShut {NoStop}%
\bibitem [{\citenamefont {Alavi-Harati}\ \emph {et~al.}(2003)\citenamefont
  {Alavi-Harati} \emph {et~al.}}]{AlaviHarati:2002ye}%
  \BibitemOpen
  \bibfield  {author} {\bibinfo {author} {\bibfnamefont {A.}~\bibnamefont
  {Alavi-Harati}} \emph {et~al.} (\bibinfo {collaboration} {KTeV}),\ }\href
  {\doibase 10.1103/PhysRevD.70.079904, 10.1103/PhysRevD.67.012005} {\bibfield
  {journal} {\bibinfo  {journal} {Phys. Rev.}\ }\textbf {\bibinfo {volume}
  {D67}},\ \bibinfo {pages} {012005} (\bibinfo {year} {2003})},\ \bibinfo
  {note} {[Erratum: Phys. Rev.D70,079904(2004)]},\ \Eprint
  {http://arxiv.org/abs/hep-ex/0208007} {arXiv:hep-ex/0208007 [hep-ex]}
  \BibitemShut {NoStop}%
\bibitem [{\citenamefont {Lin}\ \emph {et~al.}(2008)\citenamefont {Lin} \emph
  {et~al.}}]{Lin:2008zzaa}%
  \BibitemOpen
  \bibfield  {author} {\bibinfo {author} {\bibfnamefont {S.~W.}\ \bibnamefont
  {Lin}} \emph {et~al.} (\bibinfo {collaboration} {Belle}),\ }\href {\doibase
  10.1038/nature06827} {\bibfield  {journal} {\bibinfo  {journal} {Nature}\
  }\textbf {\bibinfo {volume} {452}},\ \bibinfo {pages} {332} (\bibinfo {year}
  {2008})}\BibitemShut {NoStop}%
\bibitem [{\citenamefont {Duh}\ \emph {et~al.}(2013)\citenamefont {Duh} \emph
  {et~al.}}]{Duh:2012ie}%
  \BibitemOpen
  \bibfield  {author} {\bibinfo {author} {\bibfnamefont {Y.~T.}\ \bibnamefont
  {Duh}} \emph {et~al.} (\bibinfo {collaboration} {Belle}),\ }\href {\doibase
  10.1103/PhysRevD.87.031103} {\bibfield  {journal} {\bibinfo  {journal} {Phys.
  Rev.}\ }\textbf {\bibinfo {volume} {D87}},\ \bibinfo {pages} {031103}
  (\bibinfo {year} {2013})},\ \Eprint {http://arxiv.org/abs/1210.1348}
  {arXiv:1210.1348 [hep-ex]} \BibitemShut {NoStop}%
\bibitem [{\citenamefont {Aubert}\ \emph {et~al.}(2007)\citenamefont {Aubert}
  \emph {et~al.}}]{Aubert:2007mj}%
  \BibitemOpen
  \bibfield  {author} {\bibinfo {author} {\bibfnamefont {B.}~\bibnamefont
  {Aubert}} \emph {et~al.} (\bibinfo {collaboration} {BaBar}),\ }\href
  {\doibase 10.1103/PhysRevLett.99.021603} {\bibfield  {journal} {\bibinfo
  {journal} {Phys. Rev. Lett.}\ }\textbf {\bibinfo {volume} {99}},\ \bibinfo
  {pages} {021603} (\bibinfo {year} {2007})},\ \Eprint
  {http://arxiv.org/abs/hep-ex/0703016} {arXiv:hep-ex/0703016 [HEP-EX]}
  \BibitemShut {NoStop}%
\bibitem [{\citenamefont {Lees}\ \emph {et~al.}(2013)\citenamefont {Lees} \emph
  {et~al.}}]{Lees:2012mma}%
  \BibitemOpen
  \bibfield  {author} {\bibinfo {author} {\bibfnamefont {J.~P.}\ \bibnamefont
  {Lees}} \emph {et~al.} (\bibinfo {collaboration} {BaBar}),\ }\href {\doibase
  10.1103/PhysRevD.87.052009} {\bibfield  {journal} {\bibinfo  {journal} {Phys.
  Rev.}\ }\textbf {\bibinfo {volume} {D87}},\ \bibinfo {pages} {052009}
  (\bibinfo {year} {2013})},\ \Eprint {http://arxiv.org/abs/1206.3525}
  {arXiv:1206.3525 [hep-ex]} \BibitemShut {NoStop}%
\bibitem [{\citenamefont {Abulencia}\ \emph {et~al.}(2006)\citenamefont
  {Abulencia} \emph {et~al.}}]{Abulencia:2006psa}%
  \BibitemOpen
  \bibfield  {author} {\bibinfo {author} {\bibfnamefont {A.}~\bibnamefont
  {Abulencia}} \emph {et~al.} (\bibinfo {collaboration} {CDF}),\ }\href
  {\doibase 10.1103/PhysRevLett.97.211802} {\bibfield  {journal} {\bibinfo
  {journal} {Phys. Rev. Lett.}\ }\textbf {\bibinfo {volume} {97}},\ \bibinfo
  {pages} {211802} (\bibinfo {year} {2006})},\ \Eprint
  {http://arxiv.org/abs/hep-ex/0607021} {arXiv:hep-ex/0607021 [hep-ex]}
  \BibitemShut {NoStop}%
\bibitem [{\citenamefont {Aaltonen}\ \emph {et~al.}(2011)\citenamefont
  {Aaltonen} \emph {et~al.}}]{Aaltonen:2011qt}%
  \BibitemOpen
  \bibfield  {author} {\bibinfo {author} {\bibfnamefont {T.}~\bibnamefont
  {Aaltonen}} \emph {et~al.} (\bibinfo {collaboration} {CDF}),\ }\href
  {\doibase 10.1103/PhysRevLett.106.181802} {\bibfield  {journal} {\bibinfo
  {journal} {Phys. Rev. Lett.}\ }\textbf {\bibinfo {volume} {106}},\ \bibinfo
  {pages} {181802} (\bibinfo {year} {2011})},\ \Eprint
  {http://arxiv.org/abs/1103.5762} {arXiv:1103.5762 [hep-ex]} \BibitemShut
  {NoStop}%
\bibitem [{\citenamefont {Buccella}\ \emph {et~al.}(1995)\citenamefont
  {Buccella}, \citenamefont {Lusignoli}, \citenamefont {Miele}, \citenamefont
  {Pugliese},\ and\ \citenamefont {Santorelli}}]{Buccella:1994nf}%
  \BibitemOpen
  \bibfield  {author} {\bibinfo {author} {\bibfnamefont {F.}~\bibnamefont
  {Buccella}}, \bibinfo {author} {\bibfnamefont {M.}~\bibnamefont {Lusignoli}},
  \bibinfo {author} {\bibfnamefont {G.}~\bibnamefont {Miele}}, \bibinfo
  {author} {\bibfnamefont {A.}~\bibnamefont {Pugliese}}, \ and\ \bibinfo
  {author} {\bibfnamefont {P.}~\bibnamefont {Santorelli}},\ }\href {\doibase
  10.1103/PhysRevD.51.3478} {\bibfield  {journal} {\bibinfo  {journal}
  {Phys.Rev.}\ }\textbf {\bibinfo {volume} {D51}},\ \bibinfo {pages} {3478}
  (\bibinfo {year} {1995})},\ \Eprint {http://arxiv.org/abs/hep-ph/9411286}
  {arXiv:hep-ph/9411286 [hep-ph]} \BibitemShut {NoStop}%
\bibitem [{\citenamefont {Bhattacharya}\ and\ \citenamefont
  {Rosner}(2010)}]{Bhattacharya:2009ps}%
  \BibitemOpen
  \bibfield  {author} {\bibinfo {author} {\bibfnamefont {B.}~\bibnamefont
  {Bhattacharya}}\ and\ \bibinfo {author} {\bibfnamefont {J.~L.}\ \bibnamefont
  {Rosner}},\ }\href {\doibase 10.1103/PhysRevD.81.014026} {\bibfield
  {journal} {\bibinfo  {journal} {Phys. Rev.}\ }\textbf {\bibinfo {volume}
  {D81}},\ \bibinfo {pages} {014026} (\bibinfo {year} {2010})},\ \Eprint
  {http://arxiv.org/abs/0911.2812} {arXiv:0911.2812 [hep-ph]} \BibitemShut
  {NoStop}%
\bibitem [{\citenamefont {Brod}\ \emph {et~al.}(2012)\citenamefont {Brod},
  \citenamefont {Kagan},\ and\ \citenamefont {Zupan}}]{Brod:2011re}%
  \BibitemOpen
  \bibfield  {author} {\bibinfo {author} {\bibfnamefont {J.}~\bibnamefont
  {Brod}}, \bibinfo {author} {\bibfnamefont {A.~L.}\ \bibnamefont {Kagan}}, \
  and\ \bibinfo {author} {\bibfnamefont {J.}~\bibnamefont {Zupan}},\ }\href
  {\doibase 10.1103/PhysRevD.86.014023} {\bibfield  {journal} {\bibinfo
  {journal} {Phys.Rev.}\ }\textbf {\bibinfo {volume} {D86}},\ \bibinfo {pages}
  {014023} (\bibinfo {year} {2012})},\ \Eprint {http://arxiv.org/abs/1111.5000}
  {arXiv:1111.5000 [hep-ph]} \BibitemShut {NoStop}%
\bibitem [{\citenamefont {Hiller}\ \emph {et~al.}(2013)\citenamefont {Hiller},
  \citenamefont {Jung},\ and\ \citenamefont {Schacht}}]{Hiller:2012xm}%
  \BibitemOpen
  \bibfield  {author} {\bibinfo {author} {\bibfnamefont {G.}~\bibnamefont
  {Hiller}}, \bibinfo {author} {\bibfnamefont {M.}~\bibnamefont {Jung}}, \ and\
  \bibinfo {author} {\bibfnamefont {S.}~\bibnamefont {Schacht}},\ }\href
  {\doibase 10.1103/PhysRevD.87.014024} {\bibfield  {journal} {\bibinfo
  {journal} {Phys.Rev.}\ }\textbf {\bibinfo {volume} {D87}},\ \bibinfo {pages}
  {014024} (\bibinfo {year} {2013})},\ \Eprint {http://arxiv.org/abs/1211.3734}
  {arXiv:1211.3734 [hep-ph]} \BibitemShut {NoStop}%
\bibitem [{\citenamefont {{M\"uller}}\ \emph
  {et~al.}(2015{\natexlab{a}})\citenamefont {{M\"uller}}, \citenamefont
  {Nierste},\ and\ \citenamefont {Schacht}}]{Muller:2015lua}%
  \BibitemOpen
  \bibfield  {author} {\bibinfo {author} {\bibfnamefont {S.}~\bibnamefont
  {{M\"uller}}}, \bibinfo {author} {\bibfnamefont {U.}~\bibnamefont {Nierste}},
  \ and\ \bibinfo {author} {\bibfnamefont {S.}~\bibnamefont {Schacht}},\ }\href
  {\doibase 10.1103/PhysRevD.92.014004} {\bibfield  {journal} {\bibinfo
  {journal} {Phys.Rev.}\ }\textbf {\bibinfo {volume} {D92}},\ \bibinfo {pages}
  {014004} (\bibinfo {year} {2015}{\natexlab{a}})},\ \Eprint
  {http://arxiv.org/abs/1503.06759} {arXiv:1503.06759 [hep-ph]} \BibitemShut
  {NoStop}%
\bibitem [{\citenamefont {Grossman}\ and\ \citenamefont
  {Nir}(2012)}]{Grossman:2011zk}%
  \BibitemOpen
  \bibfield  {author} {\bibinfo {author} {\bibfnamefont {Y.}~\bibnamefont
  {Grossman}}\ and\ \bibinfo {author} {\bibfnamefont {Y.}~\bibnamefont {Nir}},\
  }\href {\doibase 10.1007/JHEP04(2012)002} {\bibfield  {journal} {\bibinfo
  {journal} {JHEP}\ }\textbf {\bibinfo {volume} {1204}},\ \bibinfo {pages}
  {002} (\bibinfo {year} {2012})},\ \Eprint {http://arxiv.org/abs/1110.3790}
  {arXiv:1110.3790 [hep-ph]} \BibitemShut {NoStop}%
\bibitem [{\citenamefont {Gersabeck}\ \emph {et~al.}(2012)\citenamefont
  {Gersabeck}, \citenamefont {Alexander}, \citenamefont {Borghi}, \citenamefont
  {Gligorov},\ and\ \citenamefont {Parkes}}]{Gersabeck:2011xj}%
  \BibitemOpen
  \bibfield  {author} {\bibinfo {author} {\bibfnamefont {M.}~\bibnamefont
  {Gersabeck}}, \bibinfo {author} {\bibfnamefont {M.}~\bibnamefont
  {Alexander}}, \bibinfo {author} {\bibfnamefont {S.}~\bibnamefont {Borghi}},
  \bibinfo {author} {\bibfnamefont {V.~V.}\ \bibnamefont {Gligorov}}, \ and\
  \bibinfo {author} {\bibfnamefont {C.}~\bibnamefont {Parkes}},\ }\href
  {\doibase 10.1088/0954-3899/39/4/045005} {\bibfield  {journal} {\bibinfo
  {journal} {J. Phys.}\ }\textbf {\bibinfo {volume} {G39}},\ \bibinfo {pages}
  {045005} (\bibinfo {year} {2012})},\ \Eprint {http://arxiv.org/abs/1111.6515}
  {arXiv:1111.6515 [hep-ex]} \BibitemShut {NoStop}%
\bibitem [{\citenamefont {Grossman}\ \emph {et~al.}(2007)\citenamefont
  {Grossman}, \citenamefont {Kagan},\ and\ \citenamefont
  {Nir}}]{Grossman:2006jg}%
  \BibitemOpen
  \bibfield  {author} {\bibinfo {author} {\bibfnamefont {Y.}~\bibnamefont
  {Grossman}}, \bibinfo {author} {\bibfnamefont {A.~L.}\ \bibnamefont {Kagan}},
  \ and\ \bibinfo {author} {\bibfnamefont {Y.}~\bibnamefont {Nir}},\ }\href
  {\doibase 10.1103/PhysRevD.75.036008} {\bibfield  {journal} {\bibinfo
  {journal} {Phys.Rev.}\ }\textbf {\bibinfo {volume} {D75}},\ \bibinfo {pages}
  {036008} (\bibinfo {year} {2007})},\ \Eprint
  {http://arxiv.org/abs/hep-ph/0609178} {arXiv:hep-ph/0609178 [hep-ph]}
  \BibitemShut {NoStop}%
\bibitem [{\citenamefont {Amhis}\ \emph {et~al.}(2012)\citenamefont {Amhis}
  \emph {et~al.}}]{Amhis:2012bh}%
  \BibitemOpen
  \bibfield  {author} {\bibinfo {author} {\bibfnamefont {Y.}~\bibnamefont
  {Amhis}} \emph {et~al.} (\bibinfo {collaboration} {Heavy Flavor Averaging
  Group}),\ }\href@noop {} {\  (\bibinfo {year} {2012})},\ \Eprint
  {http://arxiv.org/abs/{1207.1158, and online update 30 June 2014}}
  {arXiv:{1207.1158, and online update 30 June 2014} [hep-ex]} \BibitemShut
  {NoStop}%
\bibitem [{\citenamefont {Beringer}\ \emph {et~al.}(2012)\citenamefont
  {Beringer} \emph {et~al.}}]{Beringer:1900zz}%
  \BibitemOpen
  \bibfield  {author} {\bibinfo {author} {\bibfnamefont {J.}~\bibnamefont
  {Beringer}} \emph {et~al.} (\bibinfo {collaboration} {Particle Data Group}),\
  }\href {\doibase 10.1103/PhysRevD.86.010001} {\bibfield  {journal} {\bibinfo
  {journal} {Phys.Rev.}\ }\textbf {\bibinfo {volume} {D86}},\ \bibinfo {pages}
  {010001} (\bibinfo {year} {2012})},\ \bibinfo {note} {{and 2013 partial
  update for the 2014 edition}}\BibitemShut {NoStop}%
\bibitem [{\citenamefont {Aaij}\ \emph {et~al.}(2014)\citenamefont {Aaij} \emph
  {et~al.}}]{Aaij:2014gsa}%
  \BibitemOpen
  \bibfield  {author} {\bibinfo {author} {\bibfnamefont {R.}~\bibnamefont
  {Aaij}} \emph {et~al.} (\bibinfo {collaboration} {LHCb collaboration}),\
  }\href {\doibase 10.1007/JHEP07(2014)041} {\bibfield  {journal} {\bibinfo
  {journal} {JHEP}\ }\textbf {\bibinfo {volume} {1407}},\ \bibinfo {pages}
  {041} (\bibinfo {year} {2014})},\ \Eprint {http://arxiv.org/abs/1405.2797}
  {arXiv:1405.2797 [hep-ex]} \BibitemShut {NoStop}%
\bibitem [{\citenamefont {Aaij}\ \emph {et~al.}(2015)\citenamefont {Aaij} \emph
  {et~al.}}]{Aaij:2015yda}%
  \BibitemOpen
  \bibfield  {author} {\bibinfo {author} {\bibfnamefont {R.}~\bibnamefont
  {Aaij}} \emph {et~al.} (\bibinfo {collaboration} {LHCb}),\ }\href {\doibase
  10.1007/JHEP04(2015)043} {\bibfield  {journal} {\bibinfo  {journal} {JHEP}\
  }\textbf {\bibinfo {volume} {1504}},\ \bibinfo {pages} {043} (\bibinfo {year}
  {2015})},\ \Eprint {http://arxiv.org/abs/1501.06777} {arXiv:1501.06777
  [hep-ex]} \BibitemShut {NoStop}%
\bibitem [{\citenamefont {Bonvicini}\ \emph {et~al.}(2001)\citenamefont
  {Bonvicini} \emph {et~al.}}]{Bonvicini:2000qm}%
  \BibitemOpen
  \bibfield  {author} {\bibinfo {author} {\bibfnamefont {G.}~\bibnamefont
  {Bonvicini}} \emph {et~al.} (\bibinfo {collaboration} {CLEO Collaboration}),\
  }\href {\doibase 10.1103/PhysRevD.63.071101} {\bibfield  {journal} {\bibinfo
  {journal} {Phys.Rev.}\ }\textbf {\bibinfo {volume} {D63}},\ \bibinfo {pages}
  {071101} (\bibinfo {year} {2001})},\ \Eprint
  {http://arxiv.org/abs/hep-ex/0012054} {arXiv:hep-ex/0012054 [hep-ex]}
  \BibitemShut {NoStop}%
\bibitem [{\citenamefont {{M. Alexander for the LHCb
  collaboration}}()}]{TalkAlexander:2015}%
  \BibitemOpen
  \bibfield  {author} {\bibinfo {author} {\bibnamefont {{M. Alexander for the
  LHCb collaboration}}},\ }\href@noop {} {}\bibinfo {note} {{Talk at the
  European Physical Society Conference on High Energy Physics 2015,
  22-29~July~2015, Vienna, Austria, LHCb-PAPER-2015-030,
  arXiv:1508.06087}}\BibitemShut {NoStop}%
\bibitem [{\citenamefont {{M\"uller}}\ \emph
  {et~al.}(2015{\natexlab{b}})\citenamefont {{M\"uller}}, \citenamefont
  {Nierste},\ and\ \citenamefont {Schacht}}]{Muller:2015rna}%
  \BibitemOpen
  \bibfield  {author} {\bibinfo {author} {\bibfnamefont {S.}~\bibnamefont
  {{M\"uller}}}, \bibinfo {author} {\bibfnamefont {U.}~\bibnamefont {Nierste}},
  \ and\ \bibinfo {author} {\bibfnamefont {S.}~\bibnamefont {Schacht}},\
  }\href@noop {} {\  (\bibinfo {year} {2015}{\natexlab{b}})},\ \Eprint
  {http://arxiv.org/abs/1506.04121} {arXiv:1506.04121 [hep-ph]} \BibitemShut
  {NoStop}%
\bibitem [{\citenamefont {Golden}\ and\ \citenamefont
  {Grinstein}(1989)}]{Golden:1989qx}%
  \BibitemOpen
  \bibfield  {author} {\bibinfo {author} {\bibfnamefont {M.}~\bibnamefont
  {Golden}}\ and\ \bibinfo {author} {\bibfnamefont {B.}~\bibnamefont
  {Grinstein}},\ }\href {\doibase 10.1016/0370-2693(89)90353-5} {\bibfield
  {journal} {\bibinfo  {journal} {Phys.Lett.}\ }\textbf {\bibinfo {volume}
  {B222}},\ \bibinfo {pages} {501} (\bibinfo {year} {1989})}\BibitemShut
  {NoStop}%
\bibitem [{\citenamefont {Pirtskhalava}\ and\ \citenamefont
  {Uttayarat}(2012)}]{Pirtskhalava:2011va}%
  \BibitemOpen
  \bibfield  {author} {\bibinfo {author} {\bibfnamefont {D.}~\bibnamefont
  {Pirtskhalava}}\ and\ \bibinfo {author} {\bibfnamefont {P.}~\bibnamefont
  {Uttayarat}},\ }\href {\doibase 10.1016/j.physletb.2012.04.039} {\bibfield
  {journal} {\bibinfo  {journal} {Phys.Lett.}\ }\textbf {\bibinfo {volume}
  {B712}},\ \bibinfo {pages} {81} (\bibinfo {year} {2012})},\ \Eprint
  {http://arxiv.org/abs/1112.5451} {arXiv:1112.5451 [hep-ph]} \BibitemShut
  {NoStop}%
\bibitem [{\citenamefont {Atwood}\ and\ \citenamefont
  {Soni}(2013)}]{Atwood:2012ac}%
  \BibitemOpen
  \bibfield  {author} {\bibinfo {author} {\bibfnamefont {D.}~\bibnamefont
  {Atwood}}\ and\ \bibinfo {author} {\bibfnamefont {A.}~\bibnamefont {Soni}},\
  }\href {\doibase 10.1093/ptep/ptt065} {\bibfield  {journal} {\bibinfo
  {journal} {PTEP}\ }\textbf {\bibinfo {volume} {2013}},\ \bibinfo {pages}
  {0903B05} (\bibinfo {year} {2013})},\ \Eprint
  {http://arxiv.org/abs/1211.1026} {arXiv:1211.1026 [hep-ph]} \BibitemShut
  {NoStop}%
\bibitem [{\citenamefont {Bander}\ \emph {et~al.}(1979)\citenamefont {Bander},
  \citenamefont {Silverman},\ and\ \citenamefont {Soni}}]{Bander:1979px}%
  \BibitemOpen
  \bibfield  {author} {\bibinfo {author} {\bibfnamefont {M.}~\bibnamefont
  {Bander}}, \bibinfo {author} {\bibfnamefont {D.}~\bibnamefont {Silverman}}, \
  and\ \bibinfo {author} {\bibfnamefont {A.}~\bibnamefont {Soni}},\ }\href
  {\doibase 10.1103/PhysRevLett.43.242} {\bibfield  {journal} {\bibinfo
  {journal} {Phys. Rev. Lett.}\ }\textbf {\bibinfo {volume} {43}},\ \bibinfo
  {pages} {242} (\bibinfo {year} {1979})}\BibitemShut {NoStop}%
\bibitem [{\citenamefont {Beneke}\ \emph {et~al.}(1999)\citenamefont {Beneke},
  \citenamefont {Buchalla}, \citenamefont {Neubert},\ and\ \citenamefont
  {Sachrajda}}]{Beneke:1999br}%
  \BibitemOpen
  \bibfield  {author} {\bibinfo {author} {\bibfnamefont {M.}~\bibnamefont
  {Beneke}}, \bibinfo {author} {\bibfnamefont {G.}~\bibnamefont {Buchalla}},
  \bibinfo {author} {\bibfnamefont {M.}~\bibnamefont {Neubert}}, \ and\
  \bibinfo {author} {\bibfnamefont {C.~T.}\ \bibnamefont {Sachrajda}},\ }\href
  {\doibase 10.1103/PhysRevLett.83.1914} {\bibfield  {journal} {\bibinfo
  {journal} {Phys. Rev. Lett.}\ }\textbf {\bibinfo {volume} {83}},\ \bibinfo
  {pages} {1914} (\bibinfo {year} {1999})},\ \Eprint
  {http://arxiv.org/abs/hep-ph/9905312} {arXiv:hep-ph/9905312 [hep-ph]}
  \BibitemShut {NoStop}%
\bibitem [{\citenamefont {Beneke}\ \emph {et~al.}(2001)\citenamefont {Beneke},
  \citenamefont {Buchalla}, \citenamefont {Neubert},\ and\ \citenamefont
  {Sachrajda}}]{Beneke:2001ev}%
  \BibitemOpen
  \bibfield  {author} {\bibinfo {author} {\bibfnamefont {M.}~\bibnamefont
  {Beneke}}, \bibinfo {author} {\bibfnamefont {G.}~\bibnamefont {Buchalla}},
  \bibinfo {author} {\bibfnamefont {M.}~\bibnamefont {Neubert}}, \ and\
  \bibinfo {author} {\bibfnamefont {C.~T.}\ \bibnamefont {Sachrajda}},\ }\href
  {\doibase 10.1016/S0550-3213(01)00251-6} {\bibfield  {journal} {\bibinfo
  {journal} {Nucl.Phys.}\ }\textbf {\bibinfo {volume} {B606}},\ \bibinfo
  {pages} {245} (\bibinfo {year} {2001})},\ \Eprint
  {http://arxiv.org/abs/hep-ph/0104110} {arXiv:hep-ph/0104110 [hep-ph]}
  \BibitemShut {NoStop}%
\bibitem [{\citenamefont {Beneke}\ and\ \citenamefont
  {Neubert}(2003)}]{Beneke:2003zv}%
  \BibitemOpen
  \bibfield  {author} {\bibinfo {author} {\bibfnamefont {M.}~\bibnamefont
  {Beneke}}\ and\ \bibinfo {author} {\bibfnamefont {M.}~\bibnamefont
  {Neubert}},\ }\href {\doibase 10.1016/j.nuclphysb.2003.09.026} {\bibfield
  {journal} {\bibinfo  {journal} {Nucl. Phys.}\ }\textbf {\bibinfo {volume}
  {B675}},\ \bibinfo {pages} {333} (\bibinfo {year} {2003})},\ \Eprint
  {http://arxiv.org/abs/hep-ph/0308039} {arXiv:hep-ph/0308039 [hep-ph]}
  \BibitemShut {NoStop}%
\bibitem [{\citenamefont {Beneke}\ \emph {et~al.}(2005)\citenamefont {Beneke},
  \citenamefont {Feldmann},\ and\ \citenamefont {Seidel}}]{Beneke:2004dp}%
  \BibitemOpen
  \bibfield  {author} {\bibinfo {author} {\bibfnamefont {M.}~\bibnamefont
  {Beneke}}, \bibinfo {author} {\bibfnamefont {T.}~\bibnamefont {Feldmann}}, \
  and\ \bibinfo {author} {\bibfnamefont {D.}~\bibnamefont {Seidel}},\ }\href
  {\doibase 10.1140/epjc/s2005-02181-5} {\bibfield  {journal} {\bibinfo
  {journal} {Eur. Phys. J.}\ }\textbf {\bibinfo {volume} {C41}},\ \bibinfo
  {pages} {173} (\bibinfo {year} {2005})},\ \Eprint
  {http://arxiv.org/abs/hep-ph/0412400} {arXiv:hep-ph/0412400 [hep-ph]}
  \BibitemShut {NoStop}%
\bibitem [{\citenamefont {Frings}\ \emph {et~al.}(2015)\citenamefont {Frings},
  \citenamefont {Nierste},\ and\ \citenamefont {Wiebusch}}]{Frings:2015eva}%
  \BibitemOpen
  \bibfield  {author} {\bibinfo {author} {\bibfnamefont {P.}~\bibnamefont
  {Frings}}, \bibinfo {author} {\bibfnamefont {U.}~\bibnamefont {Nierste}}, \
  and\ \bibinfo {author} {\bibfnamefont {M.}~\bibnamefont {Wiebusch}},\
  }\href@noop {} {\  (\bibinfo {year} {2015})},\ \Eprint
  {http://arxiv.org/abs/1503.00859} {arXiv:1503.00859 [hep-ph]} \BibitemShut
  {NoStop}%
\bibitem [{\citenamefont {Lenz}\ \emph {et~al.}(1997)\citenamefont {Lenz},
  \citenamefont {Nierste},\ and\ \citenamefont {Ostermaier}}]{Lenz:1997aa}%
  \BibitemOpen
  \bibfield  {author} {\bibinfo {author} {\bibfnamefont {A.}~\bibnamefont
  {Lenz}}, \bibinfo {author} {\bibfnamefont {U.}~\bibnamefont {Nierste}}, \
  and\ \bibinfo {author} {\bibfnamefont {G.}~\bibnamefont {Ostermaier}},\
  }\href {\doibase 10.1103/PhysRevD.56.7228} {\bibfield  {journal} {\bibinfo
  {journal} {Phys. Rev.}\ }\textbf {\bibinfo {volume} {D56}},\ \bibinfo {pages}
  {7228} (\bibinfo {year} {1997})},\ \Eprint
  {http://arxiv.org/abs/hep-ph/9706501} {arXiv:hep-ph/9706501 [hep-ph]}
  \BibitemShut {NoStop}%
\bibitem [{\citenamefont {'t~Hooft}(1974)}]{'tHooft:1973jz}%
  \BibitemOpen
  \bibfield  {author} {\bibinfo {author} {\bibfnamefont {G.}~\bibnamefont
  {'t~Hooft}},\ }\href {\doibase 10.1016/0550-3213(74)90154-0} {\bibfield
  {journal} {\bibinfo  {journal} {Nucl.Phys.}\ }\textbf {\bibinfo {volume}
  {B72}},\ \bibinfo {pages} {461} (\bibinfo {year} {1974})}\BibitemShut
  {NoStop}%
\bibitem [{\citenamefont {Buras}\ \emph {et~al.}(1986)\citenamefont {Buras},
  \citenamefont {Gerard},\ and\ \citenamefont {Ruckl}}]{Buras:1985xv}%
  \BibitemOpen
  \bibfield  {author} {\bibinfo {author} {\bibfnamefont {A.}~\bibnamefont
  {Buras}}, \bibinfo {author} {\bibfnamefont {J.}~\bibnamefont {Gerard}}, \
  and\ \bibinfo {author} {\bibfnamefont {R.}~\bibnamefont {Ruckl}},\ }\href
  {\doibase 10.1016/0550-3213(86)90200-2} {\bibfield  {journal} {\bibinfo
  {journal} {Nucl.Phys.}\ }\textbf {\bibinfo {volume} {B268}},\ \bibinfo
  {pages} {16} (\bibinfo {year} {1986})}\BibitemShut {NoStop}%
\bibitem [{\citenamefont {Buras}\ and\ \citenamefont
  {Silvestrini}(2000)}]{Buras:1998ra}%
  \BibitemOpen
  \bibfield  {author} {\bibinfo {author} {\bibfnamefont {A.~J.}\ \bibnamefont
  {Buras}}\ and\ \bibinfo {author} {\bibfnamefont {L.}~\bibnamefont
  {Silvestrini}},\ }\href {\doibase 10.1016/S0550-3213(99)00712-9} {\bibfield
  {journal} {\bibinfo  {journal} {Nucl.Phys.}\ }\textbf {\bibinfo {volume}
  {B569}},\ \bibinfo {pages} {3} (\bibinfo {year} {2000})},\ \Eprint
  {http://arxiv.org/abs/hep-ph/9812392} {arXiv:hep-ph/9812392 [hep-ph]}
  \BibitemShut {NoStop}%
\bibitem [{\citenamefont {Wiebusch}(2013)}]{Wiebusch:2012en}%
  \BibitemOpen
  \bibfield  {author} {\bibinfo {author} {\bibfnamefont {M.}~\bibnamefont
  {Wiebusch}},\ }\href {\doibase 10.1016/j.cpc.2013.06.008} {\bibfield
  {journal} {\bibinfo  {journal} {Comput.Phys.Commun.}\ }\textbf {\bibinfo
  {volume} {184}},\ \bibinfo {pages} {2438} (\bibinfo {year} {2013})},\ \Eprint
  {http://arxiv.org/abs/1207.1446} {arXiv:1207.1446 [hep-ph]} \BibitemShut
  {NoStop}%
\bibitem [{\citenamefont {Binosi}\ and\ \citenamefont
  {Theussl}(2004)}]{Binosi:2003yf}%
  \BibitemOpen
  \bibfield  {author} {\bibinfo {author} {\bibfnamefont {D.}~\bibnamefont
  {Binosi}}\ and\ \bibinfo {author} {\bibfnamefont {L.}~\bibnamefont
  {Theussl}},\ }\href {\doibase 10.1016/j.cpc.2004.05.001} {\bibfield
  {journal} {\bibinfo  {journal} {Comput.Phys.Commun.}\ }\textbf {\bibinfo
  {volume} {161}},\ \bibinfo {pages} {76} (\bibinfo {year} {2004})},\ \Eprint
  {http://arxiv.org/abs/hep-ph/0309015} {arXiv:hep-ph/0309015 [hep-ph]}
  \BibitemShut {NoStop}%
\bibitem [{\citenamefont {Vermaseren}(1994)}]{Vermaseren:1994je}%
  \BibitemOpen
  \bibfield  {author} {\bibinfo {author} {\bibfnamefont {J.}~\bibnamefont
  {Vermaseren}},\ }\href {\doibase 10.1016/0010-4655(94)90034-5} {\bibfield
  {journal} {\bibinfo  {journal} {Comput.Phys.Commun.}\ }\textbf {\bibinfo
  {volume} {83}},\ \bibinfo {pages} {45} (\bibinfo {year} {1994})}\BibitemShut
  {NoStop}%
\end{thebibliography}%

\end{document}